\documentclass[%
 reprint,
 superscriptaddress,
nofootinbib,
 amsmath,amssymb,
 aps, prd
floatfix,
]{revtex4-1} 

\usepackage{graphicx}
\usepackage{dcolumn}
\usepackage{bm}
\usepackage{hyperref}
\hypersetup{colorlinks=true,citecolor=blue,filecolor=blue,urlcolor=blue,}

\newcommand{\sigvr}{$\sigma_{v_{||}}(r)$}
\newcommand{\lcdm}{$\Lambda$CDM}
\newcommand{\apjl}{ApJ}
\newcommand{\aj}{AJ}

\newcommand{\mnras}{MNRAS}
\newcommand{\apjs}{ApJS}
\newcommand{\aap}{A\&A}

\newcommand{\jcap}{JCAP}
\newcommand{\pasp}{PASP}

\newcommand{\leftparbox}[2]{\parbox{#1}{\begin{flushleft} #2 \end{flushleft}}}

\newcommand{\twoonesig}[4][\pbwidth]{
\begin{equation}
\left.
 \begin{aligned}
#2 \\ #3
 \end{aligned}
\ \right\} \ \ \mbox{\text{\leftparbox{#1}{68\,\%,~#4}}}
\end{equation}
}

\newcommand{\threeonesig}[5][\pbwidth]{
\begin{equation}
\left.
 \begin{aligned}
#2 \\ #3 \\ #4
 \end{aligned}
\ \right\} \ \ \mbox{\text{\leftparbox{#1}{68\,\%,~#5}}}
\end{equation}
}

\newcommand{\dataplus}{\allowbreak+}
\newcommand{\mksym}[1]{\ifmmode {\rm #1}\else #1\fi}

\begin{document}

\title{Beyond BAO: improving cosmological constraints from BOSS with measurement of the void-galaxy cross-correlation}

\author{Seshadri Nadathur}
 \affiliation{Institute of Cosmology and Gravitation, University of Portsmouth, Burnaby Road, Portsmouth, PO1 3FX, UK}
\author{Paul M. Carter}%
 \affiliation{Institute of Cosmology and Gravitation, University of Portsmouth, Burnaby Road, Portsmouth, PO1 3FX, UK}
\author{Will J. Percival}%
 \affiliation{Waterloo Centre for Astrophysics, Department of Physics and Astronomy, University of Waterloo, 200 University Ave W, Waterloo, ON N2L 3G1, Canada}
 \affiliation{Perimeter Institute for Theoretical Physics, 31 Caroline St. North, Waterloo, ON N2L 2Y5, Canada}
\author{Hans A. Winther}%
 \affiliation{Institute of Cosmology and Gravitation, University of Portsmouth, Burnaby Road, Portsmouth, PO1 3FX, UK}
\author{Julian E. Bautista}%
 \affiliation{Institute of Cosmology and Gravitation, University of Portsmouth, Burnaby Road, Portsmouth, PO1 3FX, UK}

\date{\today}

\begin{abstract}
We present a measurement of the anisotropic void-galaxy cross-correlation function in the CMASS galaxy sample of the BOSS DR12 data release. We perform a joint fit to the data for redshift space distortions (RSD) due to galaxy peculiar velocities and anisotropies due to the Alcock-Paczynski (AP) effect, for the first time using a velocity field reconstruction technique to remove the complicating effects of RSD in the void centre positions themselves. Fits to the void-galaxy function give a $1\%$ measurement of the AP parameter combination $D_A(z)H(z)/c = 0.4367\pm 0.0045$ at redshift $z=0.57$, where $D_A$ is the angular diameter distance and $H$ the Hubble parameter, exceeding the precision obtainable from baryon acoustic oscillations (BAO) by a factor of $\sim3.5$ and free of systematic errors. From voids alone we also obtain a $10\%$ measure of the growth rate, $f\sigma_8(z=0.57)=0.501\pm0.051$. The parameter degeneracies are orthogonal to those obtained from galaxy clustering. Combining void information with that from BAO and galaxy RSD in the same CMASS sample, we measure $D_A(0.57)/r_s=9.383\pm 0.077$ (at 0.8\% precision), $H(0.57)r_s=(14.05\pm 0.14)\;10^3$ kms$^{-1}$Mpc$^{-1}$ (1\%) and $f\sigma_8=0.453\pm0.022$ (4.9\%), consistent with cosmic microwave background (CMB) measurements from Planck. These represent a factor $\sim2$ improvement in precision over previous results through the inclusion of void information. Fitting a flat cosmological constant $\Lambda$CDM model to these results in combination with Planck CMB data, we find up to an $11\%$ reduction in uncertainties on $H_0$ and $\Omega_m$ compared to use of the corresponding BOSS consensus values. Constraints on extended models with non-flat geometry and a dark energy of state that differs from $w=-1$ show an even greater improvement. 
\end{abstract}

\maketitle


\section{Introduction}
\label{sec:introduction}

An important observational tool in the current age of ``precision cosmology'' is measurement of the large-scale structure of the Universe at low redshifts from spectroscopic galaxy redshift surveys. Data from these surveys allow measurement of the expansion history (via the Hubble parameter $H(z)$), geometry (via the angular diameter distance $D_A(z)$) and growth rate of structure as functions of redshift. These key quantities allow precise tests of the standard cosmological $\Lambda$ Cold Dark Matter (\lcdm) model based on a spatially flat Universe with cold dark matter and a cosmological constant dark energy, as well as of alternative models of dark energy, the possible breakdown on General Relativity on large scales, the neutrino mass scale and other important questions of modern physics.

The Alcock-Paczynski effect \cite{Alcock:1979} provides a test of isotropy by comparing the observed tangential and line-of-sight dimensions of cosmological objects or large-scale clustering features. Where these objects are expected to be isotropic in the correct cosmological model, observed anisotropies may be attributed to the conversion of observed redshifts and angles to distance coordinates using the wrong assumed fiducial cosmological model. This is because distances tangential to the line of sight scale proportional to $D_A(z)$, and radial distances scale proportional to $c/H(z)$. Equating radial and tangential scales of known isotropic objects determines the ratio $F_\mathrm{AP}=D_A(z)H(z)/c$, which we refer to as the Alcock-Paczynski parameter. 

The baryon acoustic oscillation (BAO) feature in the galaxy clustering at large scales reveals a preferred length scale, the sound horizon $r_s$ at the baryon drag epoch \cite{Blake:2003,Seo:2003}, which provides a standard ruler. Because of the dependence on $r_s$, by calibrating with data from the cosmic microwave background (CMB), galaxy survey-based measurements of the BAO scale can be tied to those from the last scattering surface. The angle-averaged measure of the BAO peak location constrains the combination $(D_A^2/H)^{1/3}r_s^{-1}$. The relative locations of the BAO peak observed along and transverse to the line-of-sight constrain $F_\mathrm{AP}$ without any $r_s$ dependence \cite{Padmanabhan:2008}.  

The peculiar velocities of galaxies caused by the growth of large-scale structure affect their measured redshifts and therefore imprint additional anisotropies in the galaxy clustering along the line-of-sight, an effect known as redshift-space distortions (RSD) \cite{Kaiser:1987}. Measurement of RSD can be used to deduce the growth rate, parameterized by $f(z)\sigma_8(z)$ \cite{Percival:2009}. The RSD and Alcock-Paczynski effects are however coupled \cite{Ballinger:1996,Matsubara:1996} and therefore difficult to separate in the broadband part of the galaxy power spectrum below the BAO scale. Nevertheless, observations of BAO and the redshift-space clustering have provided the best cosmological measurements of $f(z)\sigma_8(z)$, $D_A(z)$ and $H(z)$ from several galaxy surveys so far \cite{Peacock:2001,Eisenstein:2005,Guzzo:2008,Blake:2012,Beutler:2012,Ross:2015,Howlett:2015,Alam:2017}.

The anisotropic galaxy distribution around cosmic voids can also be used as an Alcock-Paczynski test \cite{Lavaux:2012}, and to measure the growth rate \cite{Paz:2013,Hamaus:2015,Hamaus:2016,Cai:2016a,Hamaus:2017a,Achitouv:2017a,Hawken:2017,Nadathur:2019a,Nadathur:2019b,Correa:2019,Achitouv:2019}. As voids are regions of low density, if the void is large enough to dominate over environmental effects from nearby structures, the average galaxy velocity in the void vicinity is directed outwards from the void centre. This causes a distortion of the cross-correlation of void centres and galaxy positions determined in redshift-space, comprising a stretching of the underdense void interior along the line of sight, and a squashing effect at the overdense void walls. In contrast to galaxy cluster regions, the density contrasts within large voids are milder, so the coupling between galaxy velocities and the mass distribution is adequately described by linear dynamics \cite{Hamaus:2014a,Nadathur:2019a}.

In this paper we present a joint analysis of the RSD and Alcock-Paczynski effects in the void-galaxy cross-correlation measured in the CMASS galaxy catalogue in the twelfth data release (DR12; \cite{Alam-DR11&12:2015}) of the Baryon Oscillation Spectroscopic Survey (BOSS; \cite{Dawson:2013}) of the Sloan Digital Sky Survey III (SDSS-III; \cite{Eisenstein-BOSS:2011}), which is currently the premier spectroscopic galaxy dataset for large-scale structure analyses. The distribution of galaxies around voids in various datasets has previously been used to perform void Alcock-Paczynski tests without accounting for RSD \cite{Sutter:2012tf,Mao:2017b}, and void RSD analyses without accounting for the Alcock-Paczynski effect \cite{Paz:2013,Hamaus:2017a,Achitouv:2017a,Hawken:2017,Achitouv:2019}. Ref.~\cite{Hamaus:2016} performed a joint RSD-AP analysis using an earlier BOSS data release than that used in this work (see also \cite{Correa:2019} for a discussion of joint analyses using simulation data). 

We perform our joint analysis using a multipole decomposition of the void-galaxy correlation in configuration space. Recent theoretical advances in modelling these multipoles \cite{Cai:2016a,Nadathur:2019a,Nadathur:2019b} have been shown to provide an accurate description of the cross-correlation in redshift space on all scales. The description of these multipole moments uses linear perturbation theory dynamics and a convolution due to dispersion in galaxy velocities around the coherent mean outflow \cite{Nadathur:2019a}. However, a significant complication has been noted, arising from additional unmodelled RSD terms due to the motion of void centres themselves \cite{Chuang:2017,Nadathur:2019a,Nadathur:2019b}. This work is the first to make use of the large-scale velocity field reconstruction technique to remove the effects of void centre motions as proposed in Ref. \cite{Nadathur:2019b}. 

After finding void centres in the post-reconstruction density field, we consider the unreconstructed redshift-space galaxy positions around these locations. The RSD model applied to describe distortions due to galaxy velocities is sufficiently accurate that any residual observed anisotropies can be used to test the Alcock-Paczynski effect. Unlike BAO, the intrinsic size of voids cannot be easily modelled, so we cannot use them as an absolute ruler. However, assuming statistical isotropy their average shape should be spherically symmetric, so they can be used to measure $F_\mathrm{AP}$. For this parameter we show that they provide a $1\%$ constraint, tighter than that obtained from BAO by a factor of $\sim3.5$.

As a result, we find that measurement of the void-galaxy correlation provides highly complementary information to that available from standard galaxy clustering techniques, with likelihoods spanning orthogonal degeneracy directions in parameter space. We show that combining this information with that from BAO and galaxy RSD techniques applied to the same CMASS data sample \cite{Gil-Marin:2016a,Cuesta:2016,Gil-Marin:2016b} leads to a factor $\sim2$ gain in precision in the measurement of $f\sigma_8(z)$, $D_A(z)$ and $H(z)$ over the previous best results, equivalent to a dramatic increase in the effective survey volume but without requiring any new data or observational time. We explore the consequences of this for cosmological model parameter determination in combination with CMB data from Planck \cite{Planck:2015params} and show that the addition of void information from CMASS leads to significant improvement in model constraints over those obtained from the final BOSS DR12 results \cite{Alam:2017}, both in the base \lcdm~model and extensions of it. Our measurements are fully consistent with the standard \lcdm~model.

The rest of this paper is organized as follows. Section \ref{sec:data} presents the details of the BOSS galaxy data, as well as the simulations and mock catalogues used for calibration of our method, estimating covariances and testing for systematics. Section \ref{sec:methodology} describes the methods used for velocity field reconstruction and void-finding to create the void catalogue used in the measurement. Section \ref{sec:theory} reviews the theoretical modelling. Section \ref{sec:likelihoods} describes the likelihood analysis and Section \ref{sec:results} presents the results of fitting to the void-galaxy correlation. Section \ref{sec:combination} describes the combination of void results with those from BAO and galaxy clustering to obtain joint constraints, and Section \ref{sec:cosmology} explores the effect on cosmological model parameters. Finally we conclude in Section \ref{sec:conclusion}.

\section{Data}
\label{sec:data}

\subsection{BOSS galaxy sample}
\label{subsec:BOSS}

We use data from the twelfth and final data release (DR12) of BOSS \cite{Dawson:2013}. BOSS collected optical spectra for over 1.5 million targets covering nearly 10,000 deg$^2$ of the sky in two hemispheres, referred to as the North and South galactic caps (NGC and SGC respectively). Large-scale structure catalogues were created from the BOSS dataset using target selection algorithms detailed in \cite{Reid-DR12:2016}. We use the CMASS galaxy catalogue, which is based on colour-magnitude cuts designed to select massive galaxies in a narrow range of stellar mass and in the redshift range $0.4\lesssim z\lesssim0.7$. These galaxies are biased tracers of the matter distribution, with a bias of $b\sim2$. This CMASS sample has also been used in other galaxy clustering analyses \cite{Gil-Marin:2016a,Gil-Marin:2016b,Cuesta:2016}. 

A second LOWZ catalogue is also available, designed to target luminous red galaxies up to redshift of $z\sim0.4$. The LOWZ catalogue has a significantly smaller volume than CMASS, as well as a more complex sky footprint due to vetoed regions following target selection changes during the observing run \cite{Reid-DR12:2016}. This results in a much smaller number of voids and therefore much weaker constraints from the analyses described below, so we restrict our attention to CMASS for simplicity. The final BOSS `consensus' cosmological analysis \cite{Alam:2017} used a combined data sample of both CMASS and LOWZ galaxies, which represents the optimal dataset. However, the changes in sky footprint between LOWZ and CMASS complicate the application of the void-finding algorithm described in Section \ref{subsec:void-finding} to the combined dataset; we leave a solution of this problem to future work and restrict ourselves to the CMASS sample for simplicity. In Sections \ref{sec:results} and \ref{sec:combination} we show that our analysis of CMASS alone already allows a large gain in information compared to the consensus analysis from the combined dataset.

To the CMASS galaxy catalogue we apply further redshift cuts $0.4<z<0.73$ to remove regions of low observed galaxy density at high and low redshifts. This also removes a small number of redshift failures, in which galaxies have been incorrectly assigned redshifts $z\sim0$.

For reconstruction and estimation of the correlation function we make use of random catalogues included in the public BOSS data release. These randoms capture the survey window, selection function and systematic effects, and contain $50$ times as many points as the CMASS galaxies.

\subsection{Mock galaxy catalogues}
\label{subsec:MD-Patchy}

To estimate the covariance matrix and test our methods we use a suite of 1000 mock galaxy catalogues created for the DR12 data using the MD-Patchy algorithm \cite{Kitaura-DR12mocks:2016}, and designed to match the clustering and observational systematics of the CMASS sample. These mocks were created using the {\small PATCHY} approximate simulation lightcones based on augmented Lagrangian perturbation theory (ALPT) \cite{Kitaura:2013}, with mock galaxies placed within dark matter haloes using a halo abundance matching algorithm trained \cite{Rodriguez-Torres:2015} on a reference full {\it N}-body simulation from the Big MultiDark suite \cite{Klypin:2016}. Masks are applied to mimic the survey footprint and selection effects for the CMASS data. 

For reasons of computational speed, the Patchy simulations are based on an approximate gravity solver that does not correctly capture the true gravitational evolution on all scales. In particular, second-order Lagrangian perturbation theory by itself cannot fully reproduce the true halo or galaxy velocities. The ALPT algorithm therefore adds a (quasi) virialized velocity component sampled from a Gaussian distribution with two free parameters (for details, see \cite{Kitaura-PATCHY:2014,Kitaura-DR12mocks:2016}). These parameters are adjusted empirically in order to match the redshift-space monopole and quadrupole of the galaxy clustering in the BOSS DR12 data. A close matching of these primary observables is achieved, however the possibility of residual small differences in other observables, that were not used for the matching, remains. In particular, we note that the BOSS consensus analysis uses different values of the fiducial linear bias parameter $b$ in analysis of BOSS DR12 data and the MD-Patchy mocks \cite{Alam:2017}. We return to this topic in Section \ref{subsec:systematics}.

\subsection{$N$-body mocks}
\label{subsec:N-body}

In addition to the MD-Patchy approximate mock catalogues, we also make use of mock galaxy catalogues in true $N$-body simulations. The first of these is the Big MultiDark simulation \cite{Klypin:2016}, which evolved $3840^3$ dark matter particles in a box of side $2.5\;h^{-1}$Gpc using cosmological parameters $\Omega_m=0.307$, $\Omega_b=0.0482$, $\Omega_\Lambda=0.693$, $n_s=0.96$, $\sigma_8=0.8228$ and $h=0.6777$. The particle mass resolution of the simulation is $2.359\times10^{10}\;M_{\odot}/h$, so halos of mass $\sim2\times10^{12}\;M_{\odot}/h$ are well-resolved ($\sim100$ dark matter particles or more). We use the halo catalogue from this simulation at redshift $z=0.52$ and populate it with mock galaxies using a 5-parameter halo occupation distribution (HOD) model \cite{Zheng:2007} with parameters fitted to match the observed clustering of CMASS galaxies \cite{Manera:2013}. This snapshot redshift is very close to the effective redshift of the CMASS data, $z=0.57$, and the small difference is corrected for below.

Our model for the void-galaxy cross-correlation has previously been extensively tested using this mock galaxy sample in the full cubic simulation box with periodic boundary conditions \cite{Nadathur:2019a,Nadathur:2019b}. To account for survey edge effects, we create a single `CMASS-like' mock catalogue from this cubic box. This is achieved by reshaping the box as described in Ref.~\cite{Manera:2013} and using two observer positions to create approximate mock catalogues matching both the NGC and SGC samples from the box, with minimal overlap. The angular and redshift selection functions of the galaxies in these mock samples are matched to those of the DR12 data. In the following, this mock catalogue is referred to as the BigMD mock.

In addition to this, to investigate the differences in void properties with cosmology and in particular the degeneracy of the growth rate with $\sigma_8$, we also ran five separate smaller $N$-body simulations with $\sigma_8=0.75,0.80,0.83$ and $0.90$ in cubic boxes of side $1.5\;h^{-1}$Gpc. The other cosmological parameters were taken to be the same as for the Big MultiDark simulation. We use halo catalogues within each of these simulations at redshifts $z=0$, $z=0.42$ and $z=0.66$ to create mock galaxy catalogues using the HOD prescription. The primary factors affecting the size and number of voids found in any galaxy sample are the mean galaxy number density and the amplitude of the galaxy clustering (or the linear galaxy bias) \cite{Nadathur:2015b,Nadathur:2016a}, and both of these quantities are empirically already very well constrained for the CMASS sample. In each simulation and at each redshift, we therefore adjust the parameters of the HOD model via a minimization procedure in order to match the values of $\overline{n}(z)$ and the galaxy power spectrum amplitude on linear scales to those of the CMASS sample to within 1\%. This ensures that we can isolate those differences in the void properties between these samples that are due to differences in $\sigma_8$ and the redshift evolution of the dark matter density field.

\section{Void catalogue construction}
\label{sec:methodology}

Throughout the following Section and in the rest of the paper, we assume a fiducial flat \lcdm ~cosmological model with $\Omega_m=0.308$ and $h=0.6736$ in order to convert observed galaxy redshifts to radial distances. Possible differences from this fiducial model are accounted for in Section \ref{subsec:APparams}.

\subsection{Reconstructing real-space void positions}
\label{subsec:reconstruction}

As explained below in Section \ref{sec:theory}, all theoretical models of the void-galaxy correlation depend on assumptions about void properties that are not satisfied if the voids are directly identified in the redshift-space galaxy distribution. To overcome this, we use the method proposed in Ref.~\cite{Nadathur:2019b} to reconstruct the approximate real-space galaxy positions by removing the effects of large-scale velocity flows before performing the void-finding. This is closely related to BAO reconstruction \cite{Eisenstein:2007}. Here we briefly recap this procedure.

The Eulerian galaxy position $\mathbf{x}$ at time $t$ is related to its initial Lagrangian position $\mathbf{q}$ by a displacement field $\mathbf{\Psi}$: $\mathbf{x}(\mathbf{q}, t) = \mathbf{q} + \mathbf{\Psi}(\mathbf{q}, t)$. $\mathbf{\Psi}$ is determined by the equation \cite{Nusser:1994}
\begin{equation}
    \label{eq:displacement}
    \nabla\cdot\mathbf{\Psi}+\frac{f}{b}\nabla\cdot(\mathbf{\Psi}\cdot\mathbf{\hat{r}})\mathbf{\hat{r}} = -\frac{\delta_{g}}{b},
\end{equation}
where $f$ is the growth rate, $b$ is the galaxy bias, and $\delta_g$ is the observed galaxy overdensity field in redshift space. Given the full solution $\mathbf{\Psi}$ to equation \ref{eq:displacement}, the component of it due to distortions to the observed redshifts by galaxy peculiar velocities alone is $\mathbf{\Psi}_\mathrm{RSD}=-f(\mathbf{\Psi}\cdot\mathbf{\hat{r}})\mathbf{\hat{r}}$. Therefore by solving for $\mathbf{\Psi}_\mathrm{RSD}$ and shifting individual galaxy positions by $-\mathbf{\Psi}_\mathrm{RSD}(\mathbf{x})$ we can move them from observed redshift space to real space and obtain the (approximate) real-space galaxy distribution.

Note that the final step differs from that used in post-reconstruction BAO analyses, in which galaxy positions are shifted by the full $-\mathbf{\Psi}$ rather than only its RSD component. This is because BAO analyses seek to undo the effects of non-linear clustering in order to sharpen the BAO peak \cite{Eisenstein:2007}, whereas we are only interested in removing RSD. A corollary is that our reconstruction procedure is only sensitive to the value of $\beta=f/b$ rather than the parameters $f$ and $b$ individually.

Practical implementation of the reconstruction step is performed using the public code \nolinkurl{REVOLVER} (REal-space VOid Locations from surVEy Reconstruction)\footnote{\url{https://github.com/seshnadathur/Revolver}}, which also subsequently performs the void-finding step described below in Section \ref{subsec:void-finding}. \nolinkurl{REVOLVER} solves equation \ref{eq:displacement} using an iterative fast Fourier transform procedure \cite{Burden:2014,Burden:2015} operating on the density field on a grid. For numerical efficiency, the grid size is chosen to be $512^3$, as in the BOSS post-reconstruction analyses \cite{Alam:2017}. For the CMASS sample, this means that the size of each grid cell is $\sim6\,h^{-1}$Mpc. We determine the galaxy density on the grid from the redshift-space galaxy positions using a weighted cloud-in-cell interpolation scheme accounting for completeness, fibre collisions, systematic and FKP weights \cite{Feldman:1994} provided for the CMASS data, and normalize this to obtain the overdensity $\delta_g$ using the BOSS random catalogues that capture the survey mask, selection and systematic effects. This $\delta_g$ is smoothed with a Gaussian kernel of width $R_s$ in order to remove sensitivity to small-scale density modes for which the Zeldovich approximation in equation \ref{eq:displacement} breaks down. The smoothing scale $R_s=10\;h^{-1}$Mpc was previously determined to be optimal for the purpose of void-galaxy correlation measurements \cite{Nadathur:2019b}; however, we checked that taking $R_s=15\;h^{-1}$Mpc as used in the BOSS analysis \cite{Alam:2017,Vargas-Magana:2017} does not affect our results.

\begin{figure}
\centering
\includegraphics[width=0.95\linewidth]{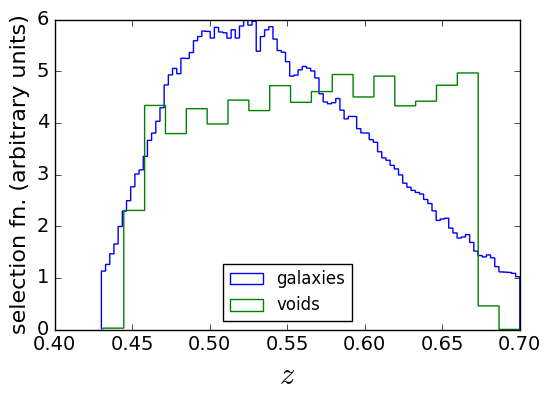}
\caption{The radial selection function for CMASS galaxies used for void-finding (blue). This selection function varies strongly with redshift due to the BOSS target selection algorithm, so CMASS is not a volume-limited survey. The green histogram shows the resultant selection function for voids in broader redshift bins (shown here for a reference catalogue obtained using $\beta=0.4$ for the reconstruction). The void distribution is relatively uniform in redshift due to the use of systematic weights to correct for the galaxy selection effects, as described in the text. Void numbers fall sharply close to upper and lower redshift extents of the survey due to the removal of edge-contamination effects.}
\label{fig:catalogues}
\end{figure}

\subsection{Void-finding}
\label{subsec:void-finding}

The reconstruction step is always performed before void-finding; void-finding is only performed on the reconstructed real-space galaxy positions. As the reconstruction results depend on $\beta$, both reconstruction and void-finding steps are repeated for each value of $\beta$ to be tested. While reconstruction is performed on all galaxies in the redshift range $0.4<z<0.73$, to avoid problems with velocity field reconstruction very close to the survey edges and from galaxies being shifted in or out of these sharp redshift cuts, void-finding is performed on a more conservative sample with redshifts $0.43<z<0.7$ after reconstruction.

Void-finding is performed by \nolinkurl{REVOLVER} using a watershed void-finding algorithm adapted from the \nolinkurl{ZOBOV} (ZOnes Bordering On Voidness) code \cite{Neyrinck:2008}. This algorithm works on the following principle. The local galaxy number density is estimated from the discrete galaxy distribution using a Voronoi Tessellation Field Estimator (VTFE) method: a Voronoi tessellation of the survey volume is performed, with each Voronoi cell being associated with a single galaxy, and the local number density is estimated from the inverse volume of each Voronoi cell normalized by the mean. This method effectively uses an adaptive scaling and is therefore more robust to Poisson noise effects. In the VTFE estimated density field the algorithm identifies local density minima as the sites of voids, and the watershed basin region around these minima determine the void extents. Following previous works \cite{Nadathur:2015b, Nadathur:2015c, Nadathur:2016a,Nadathur:2016b,Nadathur:2017a} we define each individual density basin as a distinct void, without any additional merging of neighbouring regions.

Application of the VTFE density estimation to non-uniform survey regions requires additional corrections \cite{Neyrinck:2008,Granett:2008a,Nadathur:2014a,Nadathur:2016a}. If uncorrected, the tessellation would assign arbitrarily low densities to galaxies close to unsurveyed regions. To avoid this, a set of fake `buffer' particles at 10 times the galaxy mean density is placed outside the survey edges and within holes in the mask to terminate the tessellation. Density estimates from the Voronoi cells associated with these buffer particles and any galaxies adjacent to them in the tessellation are identified and removed before the watershed stage (for details, see \cite{Nadathur:2014a,Nadathur:2016a}). VTFE density estimates are also corrected to account for the non-uniform selection function and angular completeness of the CMASS sample by applying weights \cite{Neyrinck:2008,Nadathur:2016a}. This results in a close to uniform redshift distribution of voids (Figure \ref{fig:catalogues}). The angular distribution of voids on the sky is also close to uniform. `Zones of avoidance' with no voids are present close to holes in the survey mask and at the upper and lower redshift extents of the survey, where density estimates potentially corrupted by the buffer particles have been removed. 

For each void, we take the void centre to be the centre of the largest sphere completely empty of galaxies that can be inscribed in it \cite{Nadathur:2015b}. This has been shown to be a more robust estimate of the location of the true dark matter density minimum \cite{Nadathur:2015b,Nadathur:2016a,Nadathur:2017a}, but \nolinkurl{REVOLVER} also provides alternative definitions of the void centre for comparison purposes. The effects of void centre choice were explored in Ref.~\cite{Nadathur:2019a}: choosing a different definition does not alter the validity of the RSD model described in Section \ref{sec:theory} below. The model also makes no reference to the specific void-finding algorithm, and is therefore generally applicable to any alternative void definitions as well, as long as the minimal assumptions are satisfied. However, measurement precision and the ability to resolve features in the cross-correlation multipoles can be affected by different choices for the void catalogue construction.

We determine an effective spherical radius for each void, $R_v = (3V/4\pi)^{1/3}$, where $V$ is the total volume of all Voronoi cells in the basin. For the cross-correlation measurement, we only include the largest 50\% of all obtained voids, as the RSD model is not expected to be valid for small voids where galaxy velocities are dominated by environmental effects \cite{Nadathur:2019a}. Cutting voids below the median $R_v$ is equivalent to selecting $R_v\gtrsim38\;h^{-1}$Mpc; however, the exact numerical value of the cut on $R_v$ has a slight dependence on $\beta$, as recovered void properties depend slightly on the reconstruction step \cite{Nadathur:2019b}. For $\beta=0.4$, the full void catalogue contains almost 8000 voids in the DR12 CMASS data across both NGC and SGC, so cross-correlation measurements are performed with $\sim4000$ voids. The void size distribution peaks near the median $R_v$ and drops off sharply at values $R_v\gtrsim60\;h^{-1}$Mpc, so the mean effective radius of all voids included in the cross-correlation measurements is $\simeq52\;h^{-1}$Mpc.

The same procedure, including both reconstruction and void-finding steps, is applied to each of the MD-Patchy mocks to obtain mock void catalogues, and to the mock galaxy samples in the $N$-body simulations described in Section \ref{subsec:N-body}.

\section{Theory}
\label{sec:theory}

\begin{figure}
\centering
\includegraphics[width=0.95\linewidth]{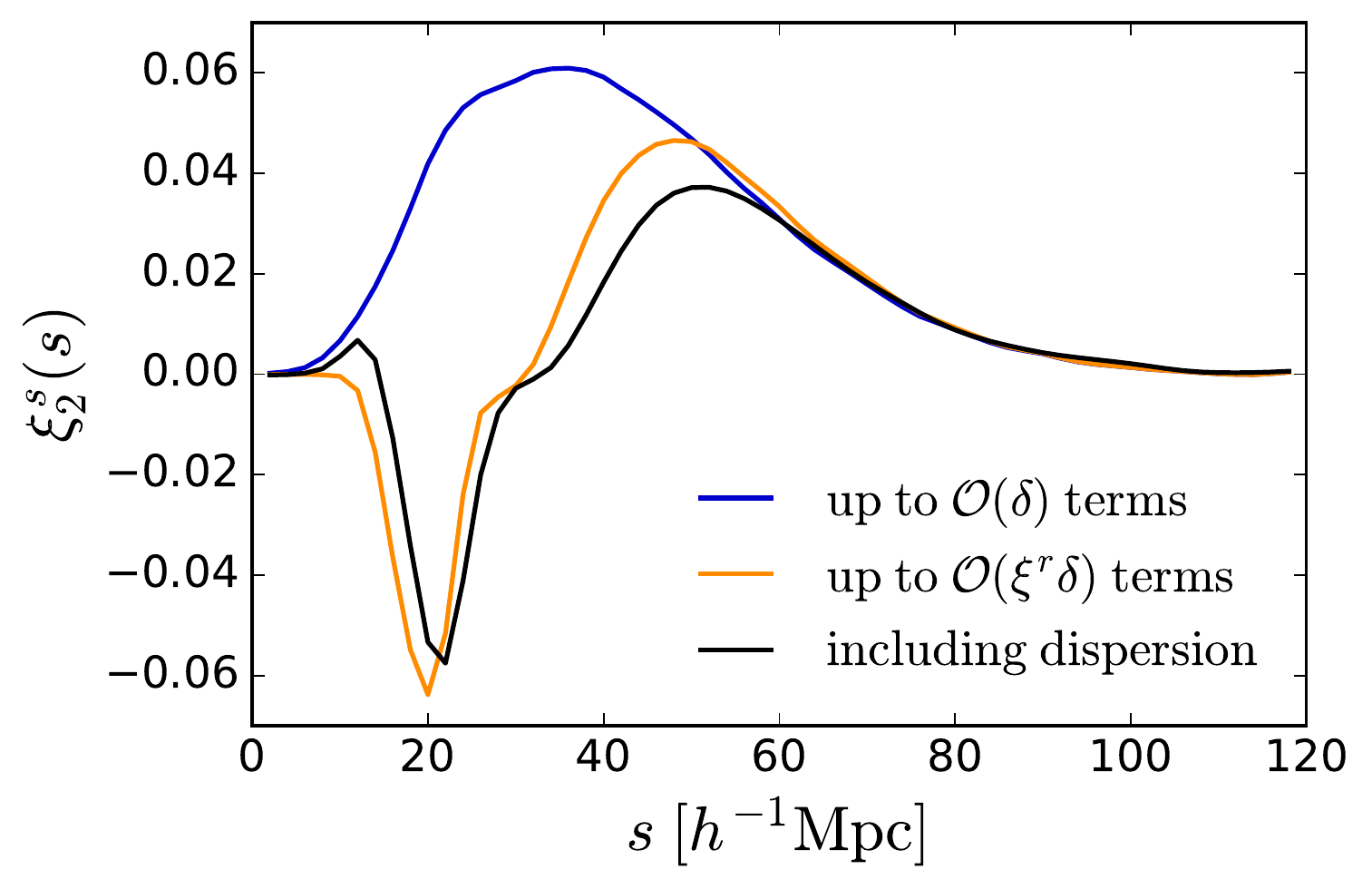}
\caption{Demonstration of the importance of using the correct theory model. We show the model quadrupole of the redshift-space void-galaxy correlation function, $\xi^s_2(s)$, calculated from equation \ref{eq:basic xi} when the expansion is truncated at terms of $\mathcal{O}(\delta)$ (blue) and at $\mathcal{O}(\xi^r\delta)$ (orange). The large correction shows that the series is not converged at $\mathcal{O}(\delta)$. The black curve shows the quadrupole when including velocity dispersion effects, equation \ref{eq:theory}, which gives corrections important for precision fits. All theory curves are calculated for representative inputs for CMASS voids, calibrated on simulation as in Section \ref{subsec:calibration}, and for $\alpha_\perp=\alpha_{||}=1$.}
\label{fig:theoryseries}
\end{figure}

\subsection{The void-galaxy cross-correlation}
\label{subsec:model}

We use the theoretical model for RSD in the void-galaxy cross-correlation developed in Refs. \cite{Nadathur:2019a,Nadathur:2019b} and briefly reviewed below. We denote the cross-correlation with galaxy positions in real space by $\xi^r$ and in redshift-space by $\xi^s$. The real- and redshift-space void-galaxy separation vectors are denoted $\bf{r}$ and $\bf{s}$ respectively. We make the crucial assumption that the mapping $\mathbf{r}\rightarrow\mathbf{s}$ from real to redshift space depends only on the galaxy velocities $\mathbf{v}_g$, i.e. that the void centre positions do not move, $\mathbf{V}_v=0$. This gives
\begin{equation}
    \label{eq:coords}
    \mathbf{s} = \mathbf{r} + \frac{\mathbf{v}_g\cdot\hat{\mathbf{X}}}{aH}\hat{\mathbf{X}}\,,
\end{equation}
where $\bf{X}$ is the vector from the observer to the void centre, $a=1/(1+z)$ and $H=H(z)$ is the Hubble rate at redshift $z$.

As has been previously noted \cite{Nadathur:2019b}, this assumption of stationary void centre positions is common to \emph{all} current theoretical models of the void-galaxy correlation in redshift space, but it is known to be invalid if voids are identified directly using redshift-space galaxy positions \cite{Chuang:2017,Nadathur:2019a,Nadathur:2019b,Massara:2018}. This problem necessitates the use of the reconstruction method described above in Section \ref{subsec:reconstruction}, which effectively recovers the real-space void positions.

Under the further assumptions that galaxy velocities around the void centre are determined by linear dynamics, depend only on the void itself rather than other structures in the environment, and show spherical symmetry around the void centre, we may write \cite{Peebles:1980,Hamaus:2014a,Cai:2016a,Nadathur:2019a,Nadathur:2019b}
\begin{equation}
    \label{eq:vg}
    \mathbf{v}_g = v_r(r)\hat{\mathbf{r}} = -\frac{1}{3}faH\Delta(r)r\hat{\mathbf{r}}\,,
\end{equation}
where $\Delta(r)$ is the average mass density contrast within radius $r$ of the void centre,
\begin{equation}
\label{eq:Delta defn}
\Delta(r)\equiv\frac{3}{r^3}\int_0^r \delta(y)y^2 dy\,,
\end{equation}
with $\delta(r)$ the (isotropic) average mass density profile of the void. $f=\mathrm{d}\ln D/\mathrm{d}\ln a$, with $D$ the growth factor and $a$ the scale factor, is the linear growth rate of density perturbations. We note in passing that if the reconstruction step to remove void centre motions described in Section \ref{subsec:reconstruction} is not performed, the assumption of spherical symmetry of the velocity profiles is also invalid, as voids are preferentially found in regions with larger outflow velocities along the line-of-sight direction \cite{Nadathur:2019b}.

The conservation of void-galaxy pairs, together with equations \ref{eq:coords}, \ref{eq:vg} and \ref{eq:Delta defn}, gives the relation \cite{Nadathur:2019a}
\begin{equation}
  \label{eq:xis invJ}
  1+\xi^s(\mathbf{s}) = \left(1+\xi^r(r)\right)\left[ 1 - \frac{f\Delta(r)}{3} - f\mu^2\left(\delta(r)-\Delta(r)\right)\right]^{-1}\,,
\end{equation}
where $\mu\equiv \frac{\mathbf{X}\cdot\mathbf{r}}{|\mathbf{X}||\mathbf{r}|}$ is the cosine of the angle between the void-galaxy separation vector and the line-of-sight direction. A series expansion of the square brackets on the RHS gives terms of order $\mathcal{O}(\xi^r)$, $\mathcal{O}(\delta)$, $\mathcal{O}(\xi^r\delta)$, $\mathcal{O}(\delta^2)$ and so on. Several previous works on the void-galaxy correlation \cite{Cai:2016a,Hamaus:2017a,Achitouv:2019} have used a model equivalent to truncating the expansion of equation \ref{eq:xis invJ} at $\mathcal{O}(\delta)$, but this in general severely limits the range of scales over which the model is valid, due to large corrections within the void radius when the $\mathcal{O}(\xi^r\delta)$ terms are included \cite{Nadathur:2019a}. We show this explicitly in Figure \ref{fig:theoryseries}. For the voids considered in this work, $\mathcal{O}(\delta^2)$ terms are always negligible.

Expanding equation \ref{eq:xis invJ} to the correct order gives the expression
\begin{multline}
\label{eq:basic xi}
\xi^{s,\mathrm{base}}\left(s,\mu\right) = \xi^r(r)+\frac{f}{3}\Delta(r)\left(1+\xi^r(r)\right) \\ 
+f\mu^2\left[\delta(r)-\Delta(r)\right]\left(1+\xi^r(r)\right) \,,
\end{multline}
for the `base' theory model for $\xi^s$. This base model is extended to account for dispersion in line-of-sight galaxy velocities $v_{||}$ via the convolution
\begin{widetext}
\begin{equation}
    \label{eq:theory}
    1 + \xi^{s,\mathrm{th}}(s,\mu) = \int_{-\infty}^\infty \frac{\left(1 + \xi^{s,\mathrm{base}}(s_\perp,s_{||}-v_{||}/aH)\right)}{\sqrt{2\pi}\sigma_{v_{||}}(r)}\exp\left(-\frac{v_{||}^2}{2\sigma_{v_{||}}^2(r)}\right)\,dv_{||}\,,
\end{equation}
\end{widetext}
where $s_{||}=s\mu$ and $s_\perp=s\sqrt{1-\mu^2}$ are the components of the vector $\mathbf{s}$ parallel and perpendicular to the line of sight respectively, $r$ is the real-space void-galaxy separation distance
\begin{equation}
    r=\sqrt{s_\perp^2 + \left(s_{||}-\frac{(v_{||}-v_r(r)\mu)}{aH}\right)^2}\,,
\end{equation}
and $\sigma_{v_{||}}(r)$ is a position-dependent dispersion whose form is to be specified.

It is worth considering the physical interpretation of the shape of the quadrupole $\xi^s_2$ shown in Figure~\ref{fig:theoryseries}. Given the sign convention for the cross-correlation, a negative quadrupole within the mean void radius corresponds to stretching of voids along the line-of-sight direction due to galaxy outflow velocities, in accordance with an intuitive RSD picture. This effect arises due to a shift in the separation distance $r\rightarrow s$ under the RSD mapping that is due to terms of $\mathcal{O}(\xi^r\delta)$ and does not have a direct analogue in the corresponding Kaiser model for the galaxy autocorrelation \cite{Nadathur:2019a}. The region of $\xi^s_2>0$ corresponding to a squashing along the line of sight coincides with the slight overdensity at the void walls \cite{Cai:2016a}.

\begin{figure*}
\centering
\includegraphics[width=0.47\linewidth]{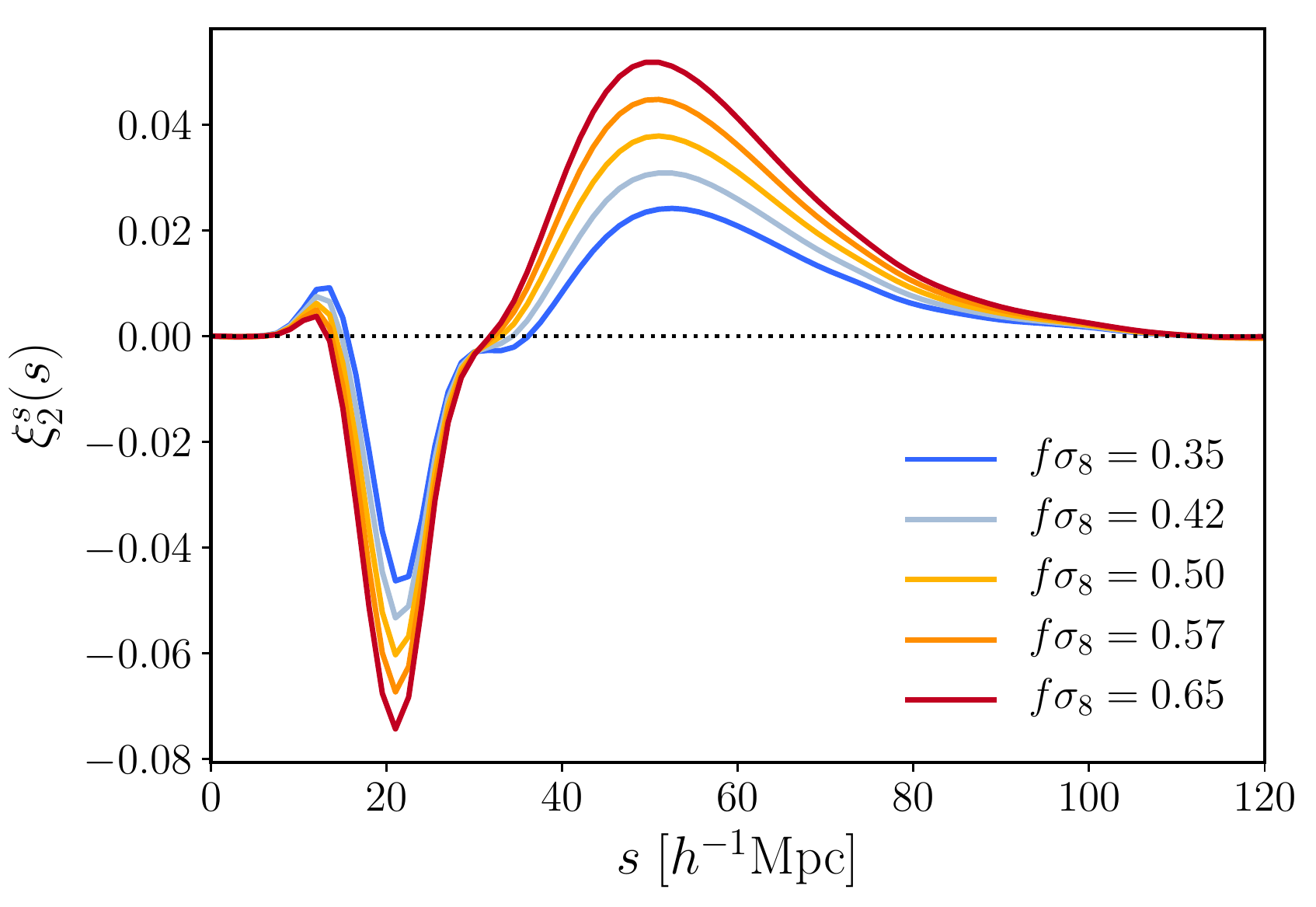}
\includegraphics[width=0.47\linewidth]{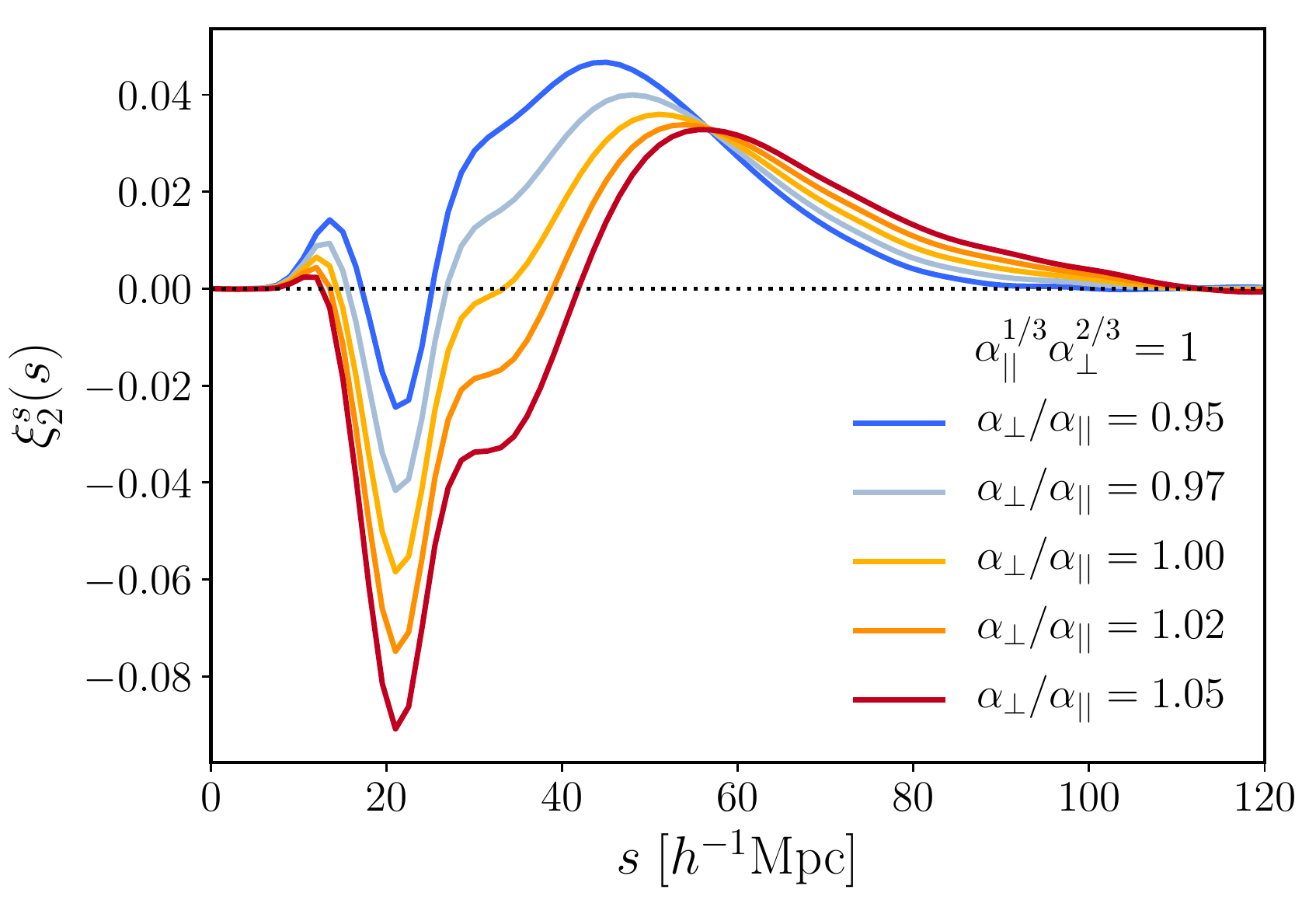}
\caption{\emph{Left}: The effect on the predicted void-galaxy quadrupole moment of changes to the assumed growth rate, parametrized by $f\sigma_8$. All other parameters are fixed to representative values for the CMASS voids as determined in Section \ref{subsec:calibration}, and $\alpha_\perp=\alpha_{||}=1$. \emph{Right}: The effect of an anisotropic rescaling of radial and transverse distances caused by assuming the wrong fiducial cosmological model, parametrized by $\alpha_\perp$ and $\alpha_{||}$. The curves show the changes as the ratio $\alpha_\perp/\alpha_{||}$ varies by up to $\pm5\%$. All other parameters are kept fixed.}
\label{fig:theorychanges}
\end{figure*}

The calculation of the full model $\xi^{s,\mathrm{th}}(s,\mu)$ requires specification of three functions in real space, namely $\xi^r(r)$, $\delta(r)$ and $\sigma_{v_{||}}(r)$, that are not known \emph{a priori}. These functions are obtained by calibration with simulation results, as described below in Section \ref{subsec:calibration}. We note that in equations \ref{eq:basic xi} and \ref{eq:theory}, the growth rate appears only as the product $f\delta(r)$ and $f\Delta(r)$. As discussed in Section \ref{subsec:calibration}, the amplitude of the void density profile $\delta(r)$ is directly proportional to $\sigma_8(z)$ and therefore the theory model depends on the degenerate combination $f(z)\sigma_8(z)$. Figure \ref{fig:theorychanges} shows the behaviour of the predicted quadrupole as $f\sigma_8$ is varied over a wide range.

For comparison with data, we decompose the correlation function into Legendre multipoles $\xi_\ell(s)$, given by
\begin{equation}
    \label{eq:Legendre}
    \xi_\ell(s)\equiv\frac{2\ell+1}{2}\int_{-1}^{1}L_\ell(\mu)\xi(s,\mu)\,\mathrm{d}\mu\,,
\end{equation}
where $L_\ell(\mu)$ is the Legendre polynomial of order $\ell$. Equations \ref{eq:basic xi} and \ref{eq:theory} mean that the only non-zero multipoles of the model are the monopole and quadrupole, $\ell=0,2$. Note also that the assumption of statistical isotropy means that for the real-space correlation $\xi^r(r)=\xi^r_0(r)$.

\subsection{Parametrizing distance scales}
\label{subsec:APparams}

 Differences between the fiducial model we have adopted for our analysis and the true cosmology will introduce additional anisotropic Alcock-Paczynski distortions by altering distances perpendicular to the line of sight and parallel to it. To parametrize this, we define the dimensionless ratios
\begin{equation}
    \label{eq:alphas}
    \alpha_\perp = \frac{D_A(z)}{D_A^\mathrm{fid}(z)},\;\;\;\alpha_{||}=\frac{H^\mathrm{fid}(z)}{H(z)}\,,
\end{equation}
to describe shifts perpendicular and parallel to the line of sight. 
To account for the fact that the 3-dimensional correlation function $\xi^s$ is measured based on void-galaxy pair separations $\mathbf{s}^\mathrm{fid}$ in the fiducial cosmological model, we rescale this function using $\alpha_\perp$ and $\alpha_{||}$ as
\begin{equation}
    \label{eq:AP s rescaling}
    \xi^s(s_\perp,s_{||}) = \xi^{s,\mathrm{fid}}\left(\alpha_\perp s^\mathrm{fid}_\perp,\alpha_{||}s^\mathrm{fid}_{||}\right)\,.
\end{equation}
In addition to this, the real-space correlation function $\xi^r(r)$ used as an input to the theory calculation is also determined from the MD-Patchy mocks in which redshifts are converted to distances at a fixed cosmology, as described in Section \ref{subsec:calibration} below. As distances in the real Universe may differ from these, we allow an additional rescaling of the real-space distances measured in the mocks, $r^\mathrm{fid}$, by integrating over angles
\begin{equation}
    \label{eq:AP r rescaling}
    r = \int_0^1 \alpha_{||}r^\mathrm{fid}\sqrt{1 + (1-\mu^2)\left(\frac{\alpha_\perp^2}{\alpha_{||}^2}-1\right)}\,\mathrm{d}\mu\,,
\end{equation}
and take $\xi^r(r) = \xi^{r,\mathrm{fid}}(r^\mathrm{fid})$ when calculating model predictions using equations \ref{eq:basic xi} and \ref{eq:theory}. This second rescaling is equivalent to asserting that the absolute void size is not known independent of cosmology. Allowing this freedom means that the model is independent of the angle-averaged parameter combination $\alpha\equiv\alpha_\perp^{2/3}\alpha_{||}^{1/3}$, in contrast to measurements of the shift of the BAO peak position where the sound horizon at the drag epoch can be calibrated from CMB observations.

On the other hand, the redshift-space void-galaxy quadrupole $\xi^s_2(s)$ is very strongly sensitive to the ratio of the Alcock-Paczynski parameters, $\epsilon\equiv\alpha_\perp/\alpha_{||}$. This is shown in the right panel of Figure \ref{fig:theorychanges}. As the RSD model described above provides a very good description of the observed $\xi^s_2(s)$ in simulations in which the fiducial cosmology is known and $\alpha_\perp=\alpha_{||}=1$ \cite{Nadathur:2019a, Nadathur:2019b}, deviations from this model prediction therefore constrain $\epsilon$. Comparison with the corresponding $f\sigma_8$ dependence in the left panel shows both why the quadrupole is much more sensitive to the ratio $\epsilon$ than to $f\sigma_8$, and why the constraints on $f\sigma_8$ and $\epsilon$ discussed in Section \ref{sec:combination} are essentially independent of each other.

\begin{figure}
\centering
\includegraphics[width=0.95\linewidth]{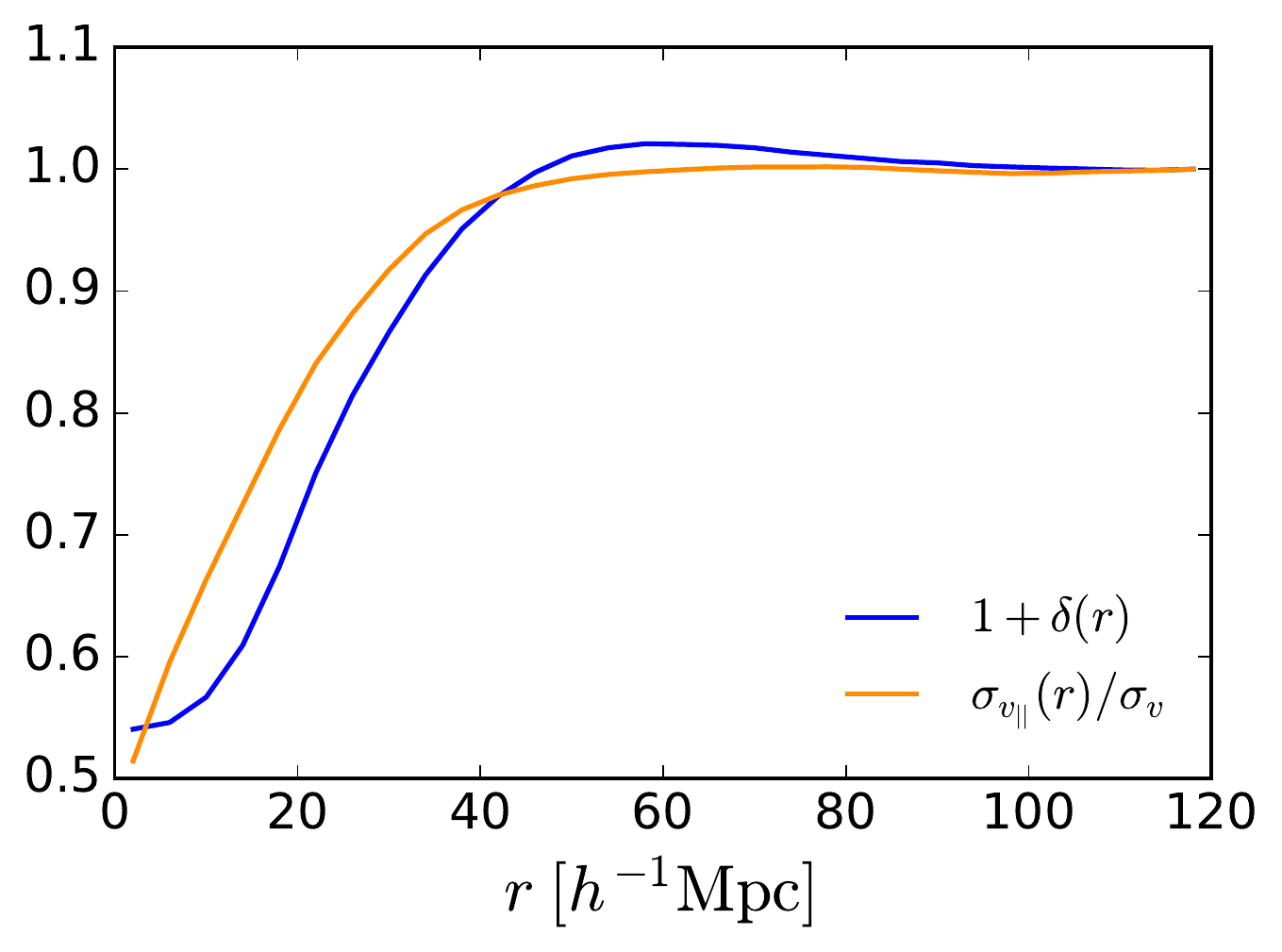}
\caption{The void-matter correlation and galaxy velocity dispersion profiles, calibrated from simulation. The fiducial mean void matter density profile $1+\delta^\mathrm{fid}(r)$ (blue) and the galaxy velocity dispersion profile $\sigma^\mathrm{fid}_{v_{||}}(r)$ (orange) measured from the custom-built mock void and galaxy catalogue in the Big MultiDark simulation at snapshot redshift $z=0.52$ and used to calibrate theoretical predictions. Both are measured as angle-averaged functions of the real-space distance from the void centre, $r$. The dispersion profile is shown normalized by its asymptotic value at large distances.}
\label{fig:normed profiles}
\end{figure}

\subsection{Calibration with simulation}
\label{subsec:calibration}

As discussed in Section \ref{sec:theory}, the full theoretical model for the observed multipoles of the redshift-space void-galaxy correlation requires the specification of three functions that are not known \emph{a priori}: $\xi^r(r)$, $\delta(r)$ and $\sigma_{v_{||}}(r)$.\footnote{Note that there are indeed three unknown functions and not two, because the use of a fixed linear bias approximation $\delta(r)=\xi^r(r)/b$ is inaccurate within voids and leads to $\sim20\%$ errors in the predicted quadrupole \cite{Nadathur:2019a}.} Here we describe how these are obtained from calibration with simulations. We note that of these three functions, the model predictions are most strongly sensitive to $\xi^r(r)$, which determines the leading order term in equation \ref{eq:basic xi}, and only very weakly dependent on the specific shape of \sigvr.

\begin{figure*}
\centering
\includegraphics[width=0.95\linewidth]{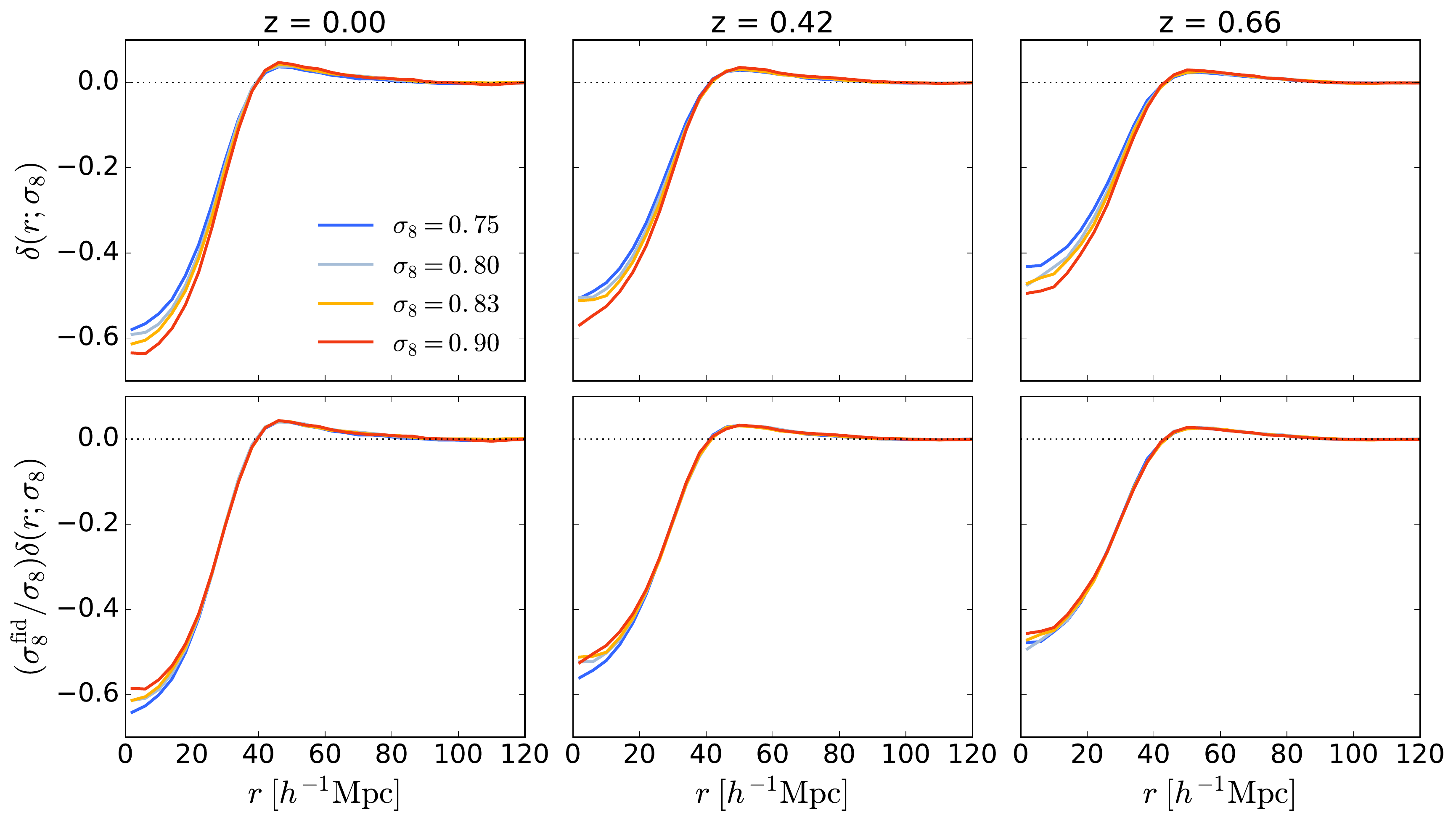}
\caption{\emph{Top row}: The mean void-matter correlation $\delta(r)$ for large voids in HOD mock galaxy catalogues in {\it N}-body simulations run with different values of the amplitude of matter density fluctuations, $\sigma_8=0.75$, 0.80, 0.83, 0.90, at redshifts $z=0$ (left column), $z=0.42$ (centre) and $z=0.66$ (right). The simulations had the same initial conditions and all other parameters were held fixed. The HOD parameters were adjusted to obtain the same mean galaxy number density and clustering amplitude for all samples, matching those of the CMASS galaxies to within 1\%. This isolates profile changes that are due to differences in cosmology and redshift only. Curves shown are for voids larger than the median in each case. \emph{Bottom row}: The same $\delta(r)$ profiles with amplitudes rescaled in proportion to the $\sigma_8$ value for each simulation, as expected in linear theory.  }
\label{fig:s8scaling}
\end{figure*}

To determine $\xi^r(r)$ we use the MD-Patchy mocks. For each of the 1000 mocks, we run the reconstruction and create the void catalogue using the \nolinkurl{REVOLVER} code exactly as done for the data sample. We then measure the real-space monopole $\xi^{r,i}_0(r)$ for the $i$th mock by cross-correlating the void centre positions with the \emph{reconstructed real-space} galaxy positions. We determine the average of these mock monopoles, $\overline{\xi^r_0}(r)$, and use this average to estimate the underlying real-space monopole:
\begin{equation}
    \label{eq:xir fiducial}
    \xi^{r,\mathrm{fid}}(r^\mathrm{fid}) = \overline{\xi^{r,\mathrm{fid}}_0}(r^\mathrm{fid})\,.
\end{equation}
Here we have added the superscript $^\mathrm{fid}$ to indicate that the mock monopoles are measured using distances calculated in the fiducial cosmological model. This fiducial monopole is rescaled using the Alcock-Paczynski parameters as described above to account for differences in cosmology. Note that as the reconstruction depends on parameter $\beta=f/b$, there is an implicit $\beta$-dependence in this quantity, $\xi^r(r)=\xi^r(r;\beta)$. We therefore recalculate $\xi^r$ at each value of $\beta$ as done for the redshift-space correlation function.

We note that $\xi^r$ could alternatively be calibrated on results from $N$-body simulations rather than the MD-Patchy mocks. The use of MD-Patchy mocks is preferable, however, as the survey systematic effects of the CMASS data have all been carefully included, and the volume increase due to the use of 1000 mocks greatly reduces measurement noise in the cross-correlation. $\xi^r$ should in general \emph{not} be determined directly from the data, as measurement errors in $\xi^r$ and $\xi^s$ could then be highly correlated, which will propagate through to the likelihood fits and manifest as an unreasonably small $\chi^2$.

The void matter density profile $\delta(r)$ and the velocity dispersion profile \sigvr ~are matched to those seen in full $N$-body simulations for accuracy, and because this information is not available in the MD-Patchy mocks. Specifically, we use the BigMD mock CMASS galaxy sample described in Section \ref{subsec:N-body}. To this mock galaxy sample we apply the \nolinkurl{REVOLVER} algorithm described in Section \ref{sec:methodology} to obtain a void catalogue. This void catalogue is cut based on the median void radius exactly as done for the data. Note that the void catalogue is created using the CMASS-like mocks incorporating the survey geometry: this is important because survey edge effects change the void size distribution in this case from that seen in cubic simulation boxes \cite{Nadathur:2016a}. 

We then we measure the fiducial stacked average dark matter density profile $\delta^\mathrm{fid}(r)$ for these voids using the simulation snapshot information. As for cross-correlation measurements with galaxies, we do not rescale distances by void size, thus effectively weighting each void equally \cite{Nadathur:2019a}. The scale $r$ is the amplitude of the real-space separation vector, which is known exactly in the simulation, and the superscript $^\mathrm{fid}$ is again used to indicate that this is the density profile in the fiducial BigMD cosmology. For the same voids, we use the velocities of the mock galaxies to measure the stacked average velocity dispersion profile $\sigma_{v_{||}}^\mathrm{fid}(r)$. Figure \ref{fig:normed profiles} shows the measured mean profiles, with $\sigma_{v_{||}}^\mathrm{fid}(r)$ shown normalized relative to its asymptotic amplitude far from the void centre, $\sigma_v\equiv\sigma_{v_{||}}^\mathrm{fid}(r\rightarrow\infty)$.

For the theory calculation, wherever $\delta(r)$ and \sigvr~are required, we use interpolating functions derived from these measured fiducial profiles. However, as the fiducial profiles have only been measured in the BigMD simulation, we also allow for variation in these profiles with differences in cosmology. This is done in two ways: changing the amplitude of the profiles, and changing their shape. 

In linear theory, the amplitude of the matter density profile $\delta(r)$ scales in proportion to the overall amplitude of the matter density perturbations parametrized by $\sigma_8(z)$. This implies that at a given redshift $z$, 
\begin{equation}
    \label{eq:s8scaling}
    \delta(r; z) = \frac{\sigma_8(z)}{\sigma_8^\mathrm{MD}(0.52)}\delta^\mathrm{fid}(r; 0.52)\,,
\end{equation}
where $\sigma_8^\mathrm{MD}(0.52) = 0.6282$ is the $\sigma_8$ value for the BigMD simulation at $z=0.52$. To verify this expected linear scaling, we used void catalogues created in the HOD galaxy mocks in each of the $N$-body simulations with different values of $\sigma_8$ (Section \ref{subsec:N-body}). The top panel of Figure \ref{fig:s8scaling} shows the measured density profiles for voids in these simulations at different redshifts, which can be seen to be appreciably different. The lower panel shows the density profiles rescaled in proportion to the simulation $\sigma_8$ values: in this case the recovered profiles are practically indistinguishable except at the void centre itself, and especially so at higher redshifts where the effects of non-linear structure growth are smaller. This justifies the use of the scaling in equation \ref{eq:s8scaling}. The dependence of the model for $\delta(r)$ on $\sigma_8$ leads to equation \ref{eq:theory} having a degeneracy between $f$ and $\sigma_8$ that matches that inherent in RSD measurements of the galaxy-galaxy clustering.

On the other hand, the amplitude of the velocity dispersion profile, while also expected to be cosmology-dependent, is not so simple to predict from theory considerations. Instead for generality we allow it to vary freely, parametrized by a free parameter $\sigma_v$, which rescales the normalized profile shown in Figure \ref{fig:normed profiles}. 

The cosmology-dependence of the shapes of the profiles $\delta(r)$ and \sigvr ~is accounted for by rescaling distances using the Alcock-Paczynski parameters, as for $\xi^r$. That is, we take $\delta(r)=\delta^\mathrm{fid}(r^\mathrm{fid})$ and $\sigma_{v_{||}}(r)=\sigma_{v_{||}}^\mathrm{fid}(r^\mathrm{fid})$, where true separation distances in the real Universe, $r$, are related to fiducial distances in the simulation, $r^\mathrm{fid}$, by equation \ref{eq:AP r rescaling}. This is again equivalent to asserting that the absolute void size cannot be known independent of an assumed cosmological model to calculate distances. In principle, the void size and thus profile shapes could also depend on other cosmological parameters, such as $\sigma_8$. However in practice this variation is very small if the mean galaxy number density and clustering amplitude are required to be close to the observed CMASS values (see, e.g., Figure \ref{fig:s8scaling}), and therefore can safely be ignored.

\section{Likelihood analysis}
\label{sec:likelihoods}

\subsection{Cross-correlation measurement}
\label{subsec:x-corr}

We measure the cross-correlation between the void centre positions and galaxies in redshift space as a function of the void-galaxy separation distance $s$ and the cosine of the angle of the separation vector to the line-of-sight direction, $\mu$, using the Landy-Szalay estimator \cite{Landy:1993}:
\begin{equation}
    \label{eq:LSestimator}
    \xi^s(s,\mu) = \frac{D_1D_2-D_1R_2-D_2R_1+R_1R_2}{R_1R_2}\,
\end{equation}
where $D_1$ refers to the void centre positions, $D_2$ to the galaxies, $R_1$ and $R_2$ to the corresponding sets of random points, and each pair $XY$ refers to the number of pairs for the given populations in the $(s,\mu)$ separation bin, normalized by the effective total number $N_XN_Y$ of such pairs. We measure $\xi^s(s,\mu)$ in 80 equal bins of $0\leq\mu\leq1$ and 30 equal radial bins $0<s<120\;h^{-1}$Mpc; note that we do not rescale distances based on apparent void size. We did not observe any significant difference between the cross-correlations measured in the NGC and SGC subsamples, so we combine pair counts in both galactic caps to estimate the final correlation functions.

The galaxy randoms $R_2$ used in equation \ref{eq:LSestimator} are provided by BOSS in the public DR12 data. These randoms contain 50 times as many points as the galaxies, and accurately represent the survey mask, completeness and selection effects. As the void distribution does not match that of the galaxies we create a new set of void randoms $R_1$. To do this we use the stacked sky distribution of voids in all 1000 MD-Patchy mock samples to determine the angular and redshift selection functions for voids. We then applied these selection functions to a Poisson point set to create a random sample $R_1$ containing 50 times as many points as the number of voids in the data. 

We decompose $\xi(s,\mu)$ into monopole $\xi^s_0(s)$ and quadrupole $\xi^s_2(s)$ components using equation \ref{eq:Legendre}, and combine these into a single data vector $\boldsymbol{\xi}^s\equiv(\xi^s_0,\xi^s_2)$ with 60 entries. Each time the reconstruction step is repeated with different values of $\beta=f/b$, the resultant void catalogue from \nolinkurl{REVOLVER} is altered so we repeat the cross-correlation measurement. Thus the data vector is a function of $\beta$, $\boldsymbol{\xi}^s=\boldsymbol{\xi}^s(\beta)$. 

For the MD-Patchy mocks, we also measure the real-space correlation monopole $\xi^r_0$ in the same way and using the same void centres, but with the galaxy positions taken in real space after reconstruction. This quantity also has an implicit dependence on $\beta$ via the reconstruction step, so is remeasured for each value of $\beta$.

\begin{figure}
\centering
\includegraphics[width=0.95\linewidth]{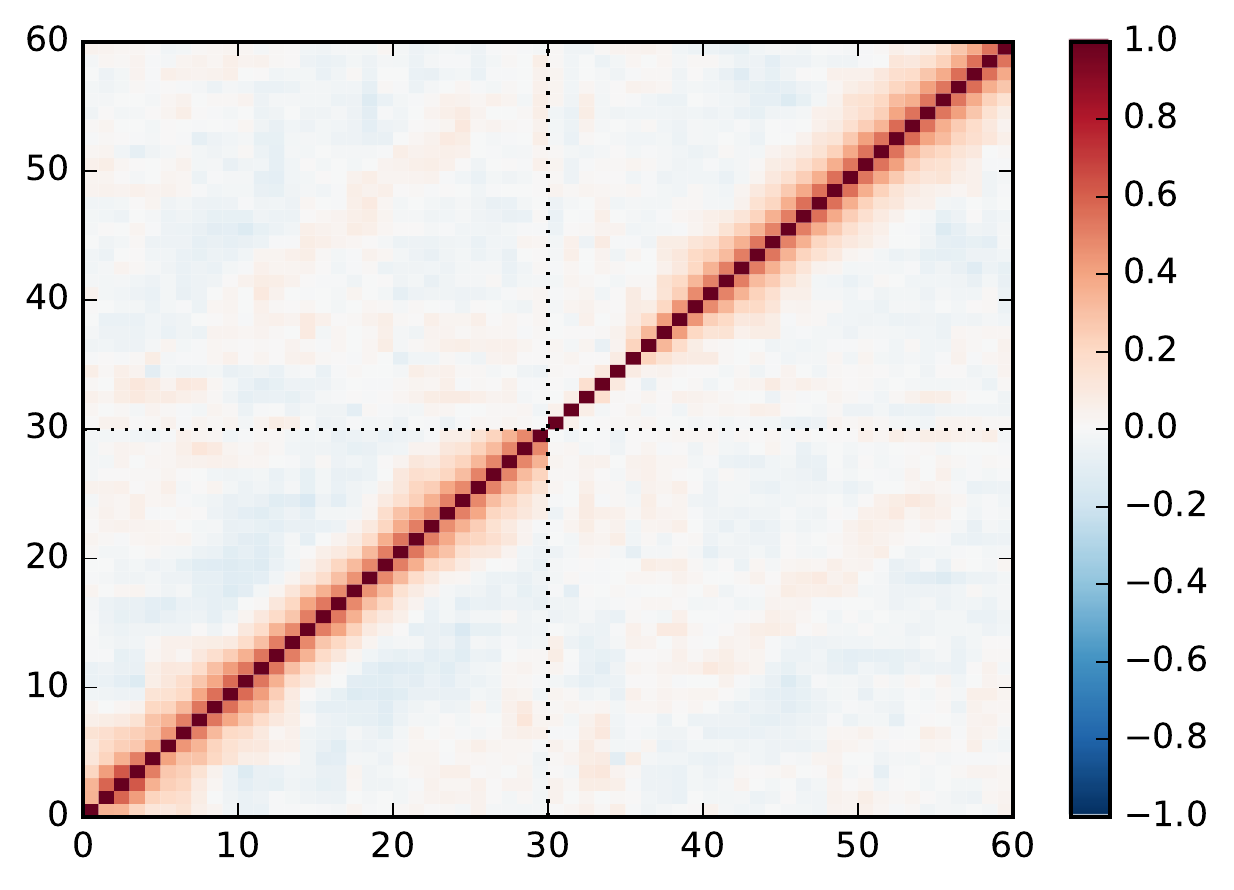}
\caption{The normalized covariance matrix for the redshift-space multipole measurement determined from 1000 MD-Patchy mocks, using a representative value $\beta=0.4$. The covariance matrix changes slightly at other values of $\beta$, which is accounted for in the likelihood determination.}
\label{fig:covmat}
\end{figure}

\subsection{Covariance matrix estimation}
\label{subsec:covmat}

We estimate the covariance matrix for our measurement using the MD-Patchy mocks. We measure the data vector $\boldsymbol{\xi}^s\equiv(\xi^s_0,\xi^s_2)$ for each of the MD-Patchy mocks, and construct the covariance matrix for the individual bin measurements as
\begin{equation}
    \label{eq:covmat}
    \mathbf{C} = \frac{1}{N-1}\sum_{k=1}^{N}\left(\boldsymbol{\xi}^k-\overline{\boldsymbol{\xi}^k}\right)\left(\boldsymbol{\xi}^k-\overline{\boldsymbol{\xi}^k}\right)\,,
\end{equation}
where $k$ is the MD-Patchy mock index, $N=1000$ and $\overline{\boldsymbol{\xi}^k}$ is the mean data vector over the mocks. As the data vector depends on $\beta$ via the reconstruction, so in principle does the covariance matrix. In practice this dependence is quite weak, but we nevertheless account for it as described below. The normalized covariance matrix for a representative value $\beta=0.4$ is shown in Figure~\ref{fig:covmat}.

\subsection{MCMC fits}
\label{subsec:fits}

The theory model for $\xi^s$ outlined in Section \ref{sec:theory} above depends explicitly on three parameters: $f\sigma_8$, $\sigma_v$, and the ratio $\alpha_\perp/\alpha_{||}$. It has a further implicit dependence on $\beta=f/b$ coming because the real-space monopole $\xi^r(r)$ used in the calculation is determined from the MD-Patchy mocks after reconstruction and thus depends on $\beta$. The data vector $\boldsymbol{\xi}^s$ depends on $\beta$, and so does the covariance matrix, $C_{ij}=C_{ij}(\beta)$. 

The full parameter space of the model is therefore 4-dimensional. If these parameters are represented as $\left(f\sigma_8,\beta,\sigma_v,\alpha_\perp/\alpha_{||}\right)$, the growth rate $f$ appears twice in different combinations. On the other hand, we are not concerned in this work with measuring the galaxy bias $b$ and so will treat it as a nuisance parameter. For convenience we therefore represent the 4 parameters as $\left(f\sigma_8,b\sigma_8,\sigma_v,\alpha_\perp/\alpha_{||}\right)$ instead, using the trivial conversion $\beta=f\sigma_8/b\sigma_8$. Of these, the two parameters of key cosmological interest are $f\sigma_8$ and $\alpha_\perp/\alpha_{||}$, with $b\sigma_8$ and $\sigma_v$ being regarded as pure nuisance parameters that are always marginalized over in quoting the final results.

We use broad uninformative priors and explore the likelihood over parameter space using Monte-Carlo Markov chains implemented via the \nolinkurl{emcee} package \cite{Foreman-Mackey:2013}. In principle, every time $f\sigma_8$ and $b\sigma_8$ are changed along the chain, the data vector and covariance matrix should be reevaluated. To make this process computationally feasible, we speed up the likelihood evaluation by using an interpolation scheme. We perform the reconstruction, void-finding and cross-correlation steps for the CMASS data and each of the MD-Patchy mocks for 30 values on a grid in the range $0.16<\beta<0.65$. We use the values of $\boldsymbol{\xi}^s$, $\xi^r(r)$ and $C_{ij}$ determined on the grid to build spline interpolations which we use to estimate $\boldsymbol{\xi}^s(\beta)$, $\xi^r(r; \beta)$ and $C_{ij}(\beta)$ at all intermediate points. We tested the accuracy of this scheme by comparing the interpolated values to those directly evaluated on a finer grid and found that in all cases the differences introduced by interpolation were much smaller than measurement errors.

As the covariance matrix is estimated from a finite number of mocks, we calculate the likelihood for the model fits to the data following the prescription of Sellentin \& Heavens \cite{Sellentin:2016}. That is, for each step in the chain we calculate
\begin{equation}
    \label{eq:chi2}
    \chi^2 = \left(\boldsymbol{\xi}^{s,\mathrm{th}}-\boldsymbol{\xi}^s\right) \mathbf{C}^{-1} \left(\boldsymbol{\xi}^{s,\mathrm{th}}-\boldsymbol{\xi}^s\right)\,,
\end{equation}
and take the log-likelihood to be
\begin{equation}
    \label{eq:lnlike}
    \ln\mathcal{L}\propto-\frac{N}{2}\ln\left(1+\frac{\chi^2}{N-1}\right)\,,
\end{equation}
where $N=1000$ is the number of MD-Patchy mocks. The chains are run until they are at least 50 times longer than the estimated autocorrelation length to ensure convergence. 

\begin{figure*}
\centering
\includegraphics[width=0.45\linewidth]{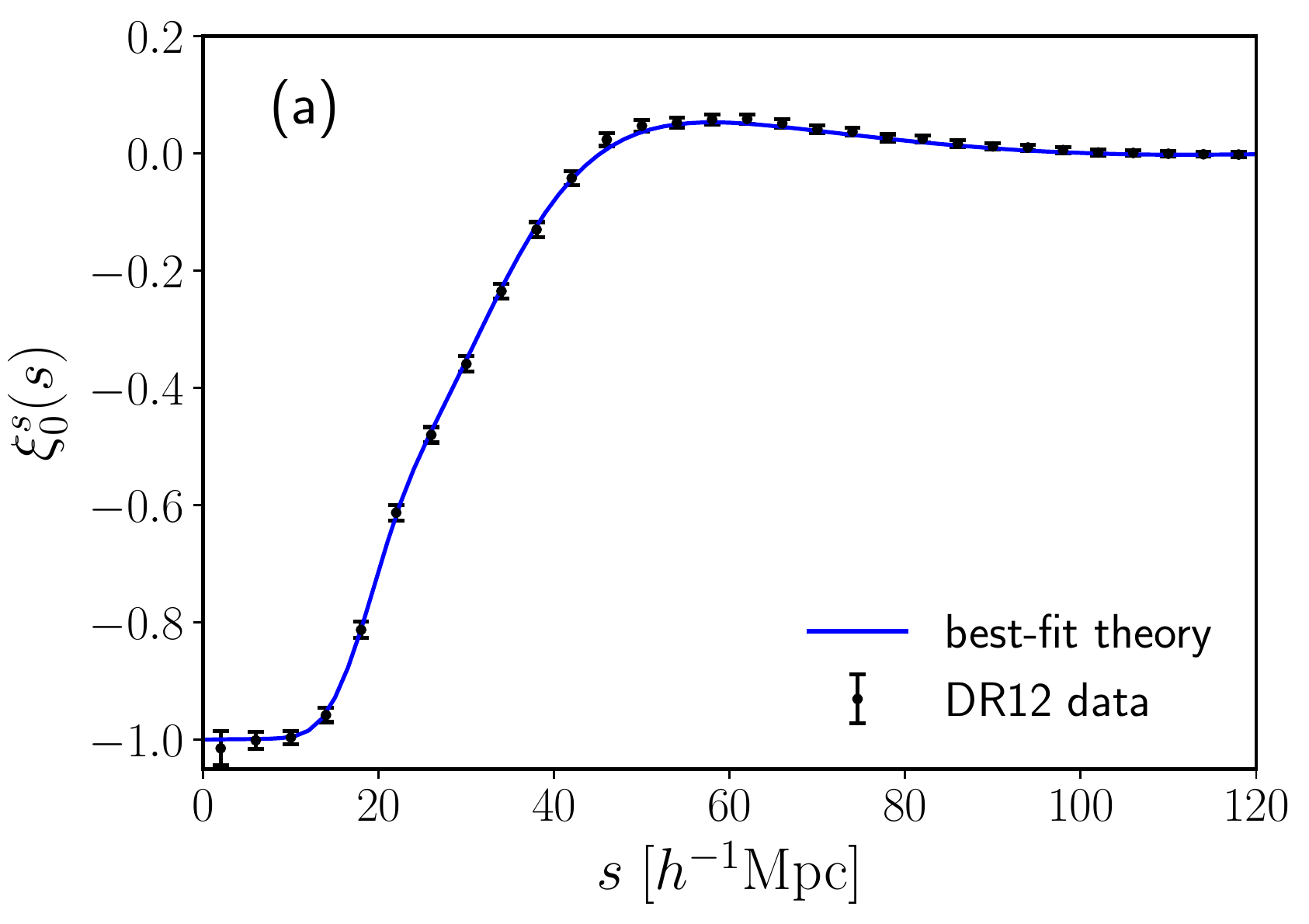}
\includegraphics[width=0.45\linewidth]{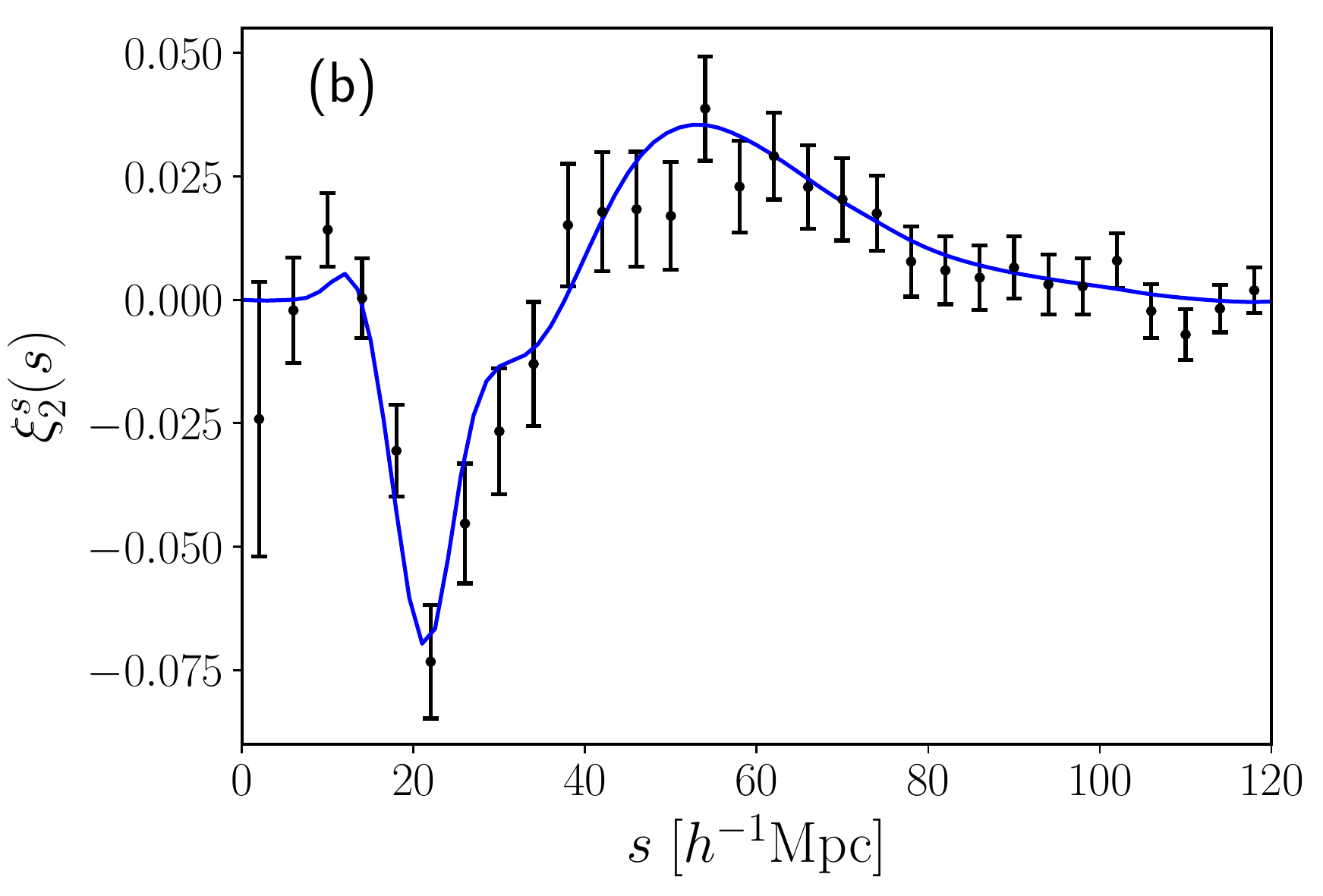}
\caption{Multipole moments of the void-galaxy cross-correlation function in redshift space from the BOSS Data Release 12 CMASS survey. The panels show (a) the angle-averaged monopole, $\xi^s_0$, and (b) the quadrupole moment, $\xi^s_2$, of the three-dimensional void-galaxy cross-correlation $\xi^s(\mathbf{s})$, as functions of the distance $s=|\mathbf{s}|$ from the void centre to the redshift-space galaxy position. Black data points show the measured values and the blue line the theoretical prediction using the linear-theory model with velocity dispersion, equation \ref{eq:theory}, for the maximum-likelihood parameter values $f\sigma_8=0.50$, $b\sigma_8=1.36$, $\sigma_v=390.7\;\mathrm{km\,s}^{-1}$, and $\alpha_\perp/\alpha_{||}=1.016$. Error bars represent the $1\sigma$ uncertainties derived from diagonal entries of the full covariance matrix estimated from 1000 MD-Patchy mocks. For panel (a), these errors are shown multiplied by a factor of 2 for visibility. }
\label{fig:fit}
\end{figure*}

\section{Results}
\label{sec:results}

\subsection{Model fits}
\label{subsec:void-only results}

\begin{figure*}
\centering
\includegraphics[width=0.9\linewidth]{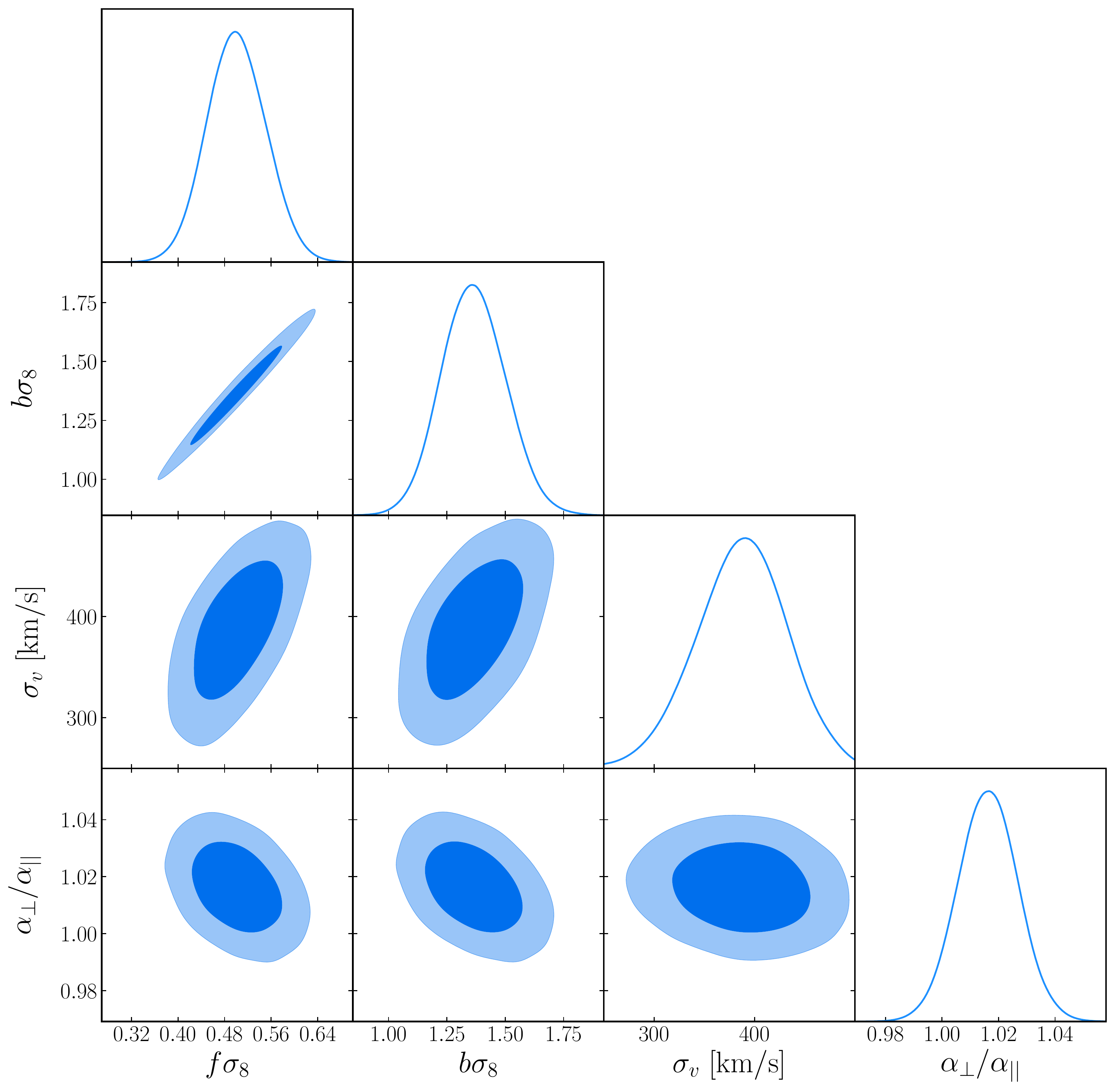}
\caption{Triangle plot showing posterior constraints on parameters of the joint RSD-AP model for the void-galaxy cross-correlation from fits to the CMASS data.}
\label{fig:triangle}
\end{figure*}

The theoretical model we have adopted provides an exceedingly good description of the measured redshift-space void-galaxy correlation. Figure \ref{fig:fit} shows a comparison between the measured multipoles and the model predictions for the maximum likelihood point in parameter space found through the MCMC analysis. The chi-squared value for this fit is $\chi^2=61.01$ for $(60-4)$ degrees of freedom, giving a reduced $\chi^2$ of 1.09. Marginalized 2D parameter constraints and degeneracies calculated from the chains are shown in Figure \ref{fig:triangle}. The mean values and marginalized 1D errors for each parameter from this analysis are: $f\sigma_8=0.501\pm 0.051$, $b\sigma_8=1.37\pm 0.14$, $\sigma_v=387^{+50}_{-40}\;\mathrm{kms}^{-1}$, $\alpha_\perp/\alpha_{||}=1.016\pm 0.011$. The derived constraint on the ratio $f/b$ is $\beta=0.37\pm0.01$.

The sensitivity of our measurement to the Alcock-Paczynski ratio $\epsilon=\alpha_\perp/\alpha_{||}$ is due to our ability to both precisely measure and model the quadrupole $\xi^s_2$, as discussed in Section \ref{subsec:APparams}. Note that the quadrupole constraints come primarily from scales in the range $15\lesssim s\lesssim60\;h^{-1}$Mpc. 

We convert the results for $\alpha_\perp/\alpha_{||}$ into constraints on the physical quantities $D_A(z)$ and $H(z)$ using equation \ref{eq:alphas}. The fits to data give a very tight, $\sim1\%$, constraint on the combination $F_\mathrm{AP}=D_A(z)H(z)/c=0.4367\pm0.0045$ at the effective redshift $z=0.57$ of the CMASS sample. This constraint is a factor of $\sim3.5$ tighter than that obtained from BAO measurement using the same CMASS data, $F_\mathrm{AP}=0.463\pm0.017$ \cite{Gil-Marin:2016a,Cuesta:2016}.\footnote{This constraint is not directly quoted in Ref. \cite{Gil-Marin:2016a}, but is easily derived from the individual constraints on $D_A/r_s$ and $Hr_s$ and the correlation coefficient provided for them.} As discussed in more detail in Section \ref{sec:combination} below, the measurement of $F_\mathrm{AP}$ is crucially not degenerate with that of the growth rate, $f\sigma_8$, which allows information to be gained by combining voids with BAO and galaxy full-shape analyses.

\subsection{Tests for systematics}
\label{subsec:systematics}

\begin{figure}
\centering
\includegraphics[width=0.95\linewidth]{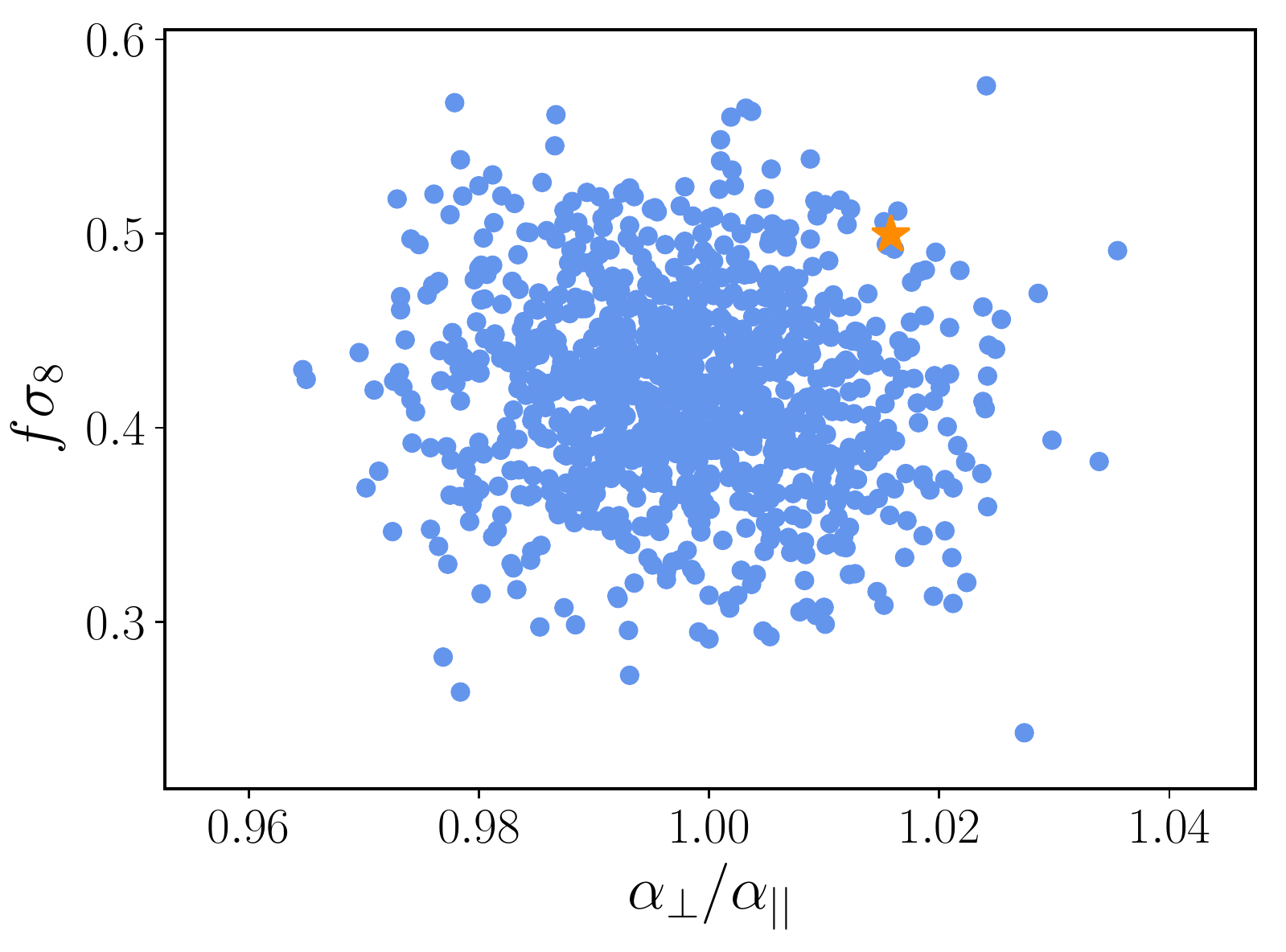}
\caption{Blue points show the maximum-likelihood parameters for fits to each of the MD-Patchy mock samples, in the $\alpha_\perp$-$\alpha_{||}$ plane (left panel) and the $f\sigma_8$-$\alpha_\perp/\alpha_{||}$ plane (right panel). The corresponding maximum-likelihood points for the DR12 CMASS data are shown by the orange stars.}
\label{fig:systematics}
\end{figure}

We use the MD-Patchy mock samples to test our analysis method for systematic errors. For each of the 1000 mocks, we apply exactly the same reconstruction, void-finding and cross-correlation measurement pipeline as used for the CMASS data. We then explore the likelihood surface to find the global maximum likelihood (ML) point for each mock, to check the distribution of these points against the statistical errors obtained from the MCMC analysis of the data. We find the ML points using a combination of local $\chi^2$ minimization and random perturbation of the starting coordinates in parameter space, in an iterative procedure implemented using the \texttt{basinhopping} algorithm \cite{Wales:1997}. 

The resulting ML values for parameters of cosmological interest are shown in Figure \ref{fig:systematics} together with the CMASS result. The scatter in the mock results in consistent with the mean values and the statistical errors derived from the MCMC analysis of CMASS data. The mean value of the Alcock-Paczynski ratio derived from the mocks is $\epsilon=0.9998\pm0.0004$ ($68\%$ c.l. error in the mean), indicative of no systematic bias in the determination of $\epsilon$ or $F_\mathrm{AP}$. As discussed below in Section \ref{sec:combination}, this is the key quantity that drives the cosmological constraints we obtain from the void analysis.

\begin{figure}
\centering
\includegraphics[width=0.95\linewidth]{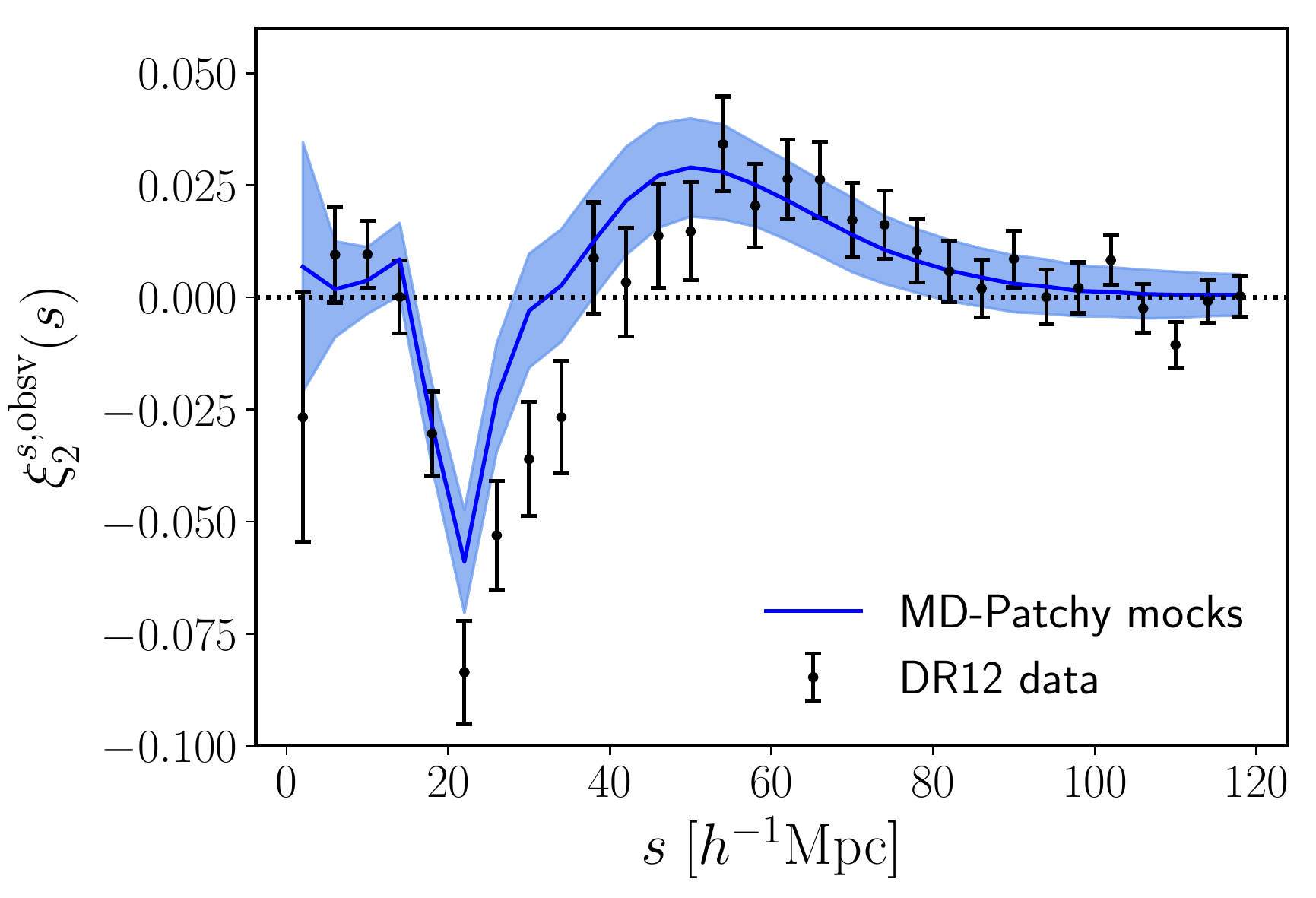}
\caption{Comparison of the measured void-galaxy quadrupole in the CMASS data (black points with error bars) and the mean of the MD-Patchy mocks (blue line), when the same fiducial growth rate $f(z=0.57)=0.78$ is used for the reconstruction step in the analysis in both cases. The shaded band shows the $68\%$ confidence level region from the distribution of the mocks. The mocks are slightly offset from the data at scales $20<s<40\;h^{-1}$Mpc, which we attribute to the approximate ALPT algorithm for assigning galaxy velocities.}
\label{fig:patchy}
\end{figure}

However, the mean value of $f\sigma_8$ obtained from the mocks is $f\sigma_8=0.445$, which is biased low compared to the fiducial value $f\sigma_8=0.48$ for the MD-Patchy cosmology. No such bias was observed when fitting the void RSD model to mocks derived from full $N$-body simulations \cite{Nadathur:2019b}. The reason for this systematic offset therefore appears to be due to a small imperfection of the MD-Patchy mocks, which do not perfectly reproduce the void-galaxy quadrupole seen in the CMASS data. In Figure \ref{fig:patchy} we show the measured quadrupole for the data and the mocks, when both are measured using the same fiducial growth rate $f(z=0.57)=0.78$ for the reconstruction step. A small but statistically significant offset is seen in the range $20<s<40\;h^{-1}$Mpc (there is no such difference in the monopole, which for simplicity we do not show). We note in this context that the MD-Patchy mocks are created using the approximate ALPT simulation method, and therefore do not automatically capture the correct dynamics on small scales. In order to match the RSD properties of the CMASS data, velocities in the MD-Patchy mocks have been tuned by adjusting two free parameters. This matching has been carried out by comparing to only two observables from the CMASS data, namely the redshift-space monopole and quadrupole of the galaxy correlation, and so observables such the void-galaxy quadrupole which were not used for matching are not guaranteed to be exactly reproduced. It is also worth pointing out that the matching in the galaxy correlation between CMASS data and the MD-Patchy mocks is only achieved at the expense of a different fiducial bias value: the final BOSS analysis uses $b_\mathrm{CMASS}=1.85$ and $b_\mathrm{Patchy}=2.10$ \cite{Alam:2017}. This suggests that the mocks cannot exactly capture the underlying physics entirely correctly.

The interpretation of this test for $f\sigma_8$ is a matter of judgement. In our opinion, the fact that the model works without bias on full $N$-body simulations and that the MD-Patchy mocks are known to be generated using approximate methods justifies attributing the offset to imperfections in how the MD-Patchy algorithm reproduces all features of the true data rather than to an intrinsic bias in the model. We therefore do not advocate adding a systematic error in $f\sigma_8$ on top of the statistical error determined from the MCMC, and do not do so in the following. However, if a very conservative analysis is desired, an alternative option is to add a systematic error of $\Delta_{f\sigma_8}=0.035$ in quadrature to the error budget. In either case, as we discuss in Section \ref{sec:combination} below, for the CMASS sample considered here addition or not of this systematic error does not affect the cosmological constraints obtained from the void analysis when combined with galaxy clustering, as these are driven primarily by the $F_\mathrm{AP}$ measurement, which is both unbiased and not degenerate with $f\sigma_8$. However, this issue may be more important for future data from DESI and Euclid, and should be investigated further using mocks developed for these surveys.

\subsection{Comparison with past work}
\label{subsec:previous}

Several other recent works have examined growth rate and Alcock-Paczynski constraints from the void-galaxy correlation. The idea of using voids to perform an Alcock-Paczynski test was first proposed in a pioneering work \cite{Lavaux:2012}, which predicted it would provide equivalent constraints to those from BAO when applied to BOSS data. Subsequently void Alcock-Paczynski tests with SDSS data have been performed in Refs. \cite{Sutter:2012tf,Sutter:2014d,Mao:2017b}. All of these works have however assumed that the anisotropy in the void-galaxy correlation is sourced entirely by the Alcock-Paczynski effect -- that is, their models did not include the additional distortion caused by RSD due to the galaxy outflow velocities around voids. For a choice of fiducial cosmological model that is relatively close to the true cosmology (and thus not grossly in conflict with constraints from other data), the dominant contribution to the void-galaxy quadrupole in redshift space in fact comes from RSD rather than the Alcock-Paczynski effect, so these studies are necessarily incomplete.

An alternative approach has been taken in several other works which fit for the RSD contribution but not the Alcock-Paczynski effect, by modelling the void-galaxy correlation using a fixed cosmological model for the distance-redshift relation \cite{Hamaus:2017a,Achitouv:2017a,Hawken:2017,Achitouv:2019}. All of these analyses assume that void centre positions do not shift in redshift space, encapsulated in the use of equation \ref{eq:coords}, which has been shown to be violated for voids identified using redshift-space galaxy positions without use of the reconstruction step applied in this work \cite{Chuang:2017,Nadathur:2019b}. Refs. \cite{Hamaus:2017a,Achitouv:2019} analyse the same DR12 CMASS data sample as in this work, using the RSD model derived by \cite{Cai:2016a}, which is equivalent to truncating the series expansion of equation \ref{eq:xis invJ} at terms of $\mathcal{O}(\delta)$. The validity of this approximation has been discussed in Section \ref{sec:theory}. They obtain constraints $\beta=0.457^{+0.056}_{-0.054}$ \cite{Hamaus:2017a} and $\beta=0.36\pm0.05$ \cite{Achitouv:2019}. 
These two papers also provide $\beta$ constraints from fits to voids in the LOWZ sample, which we have not studied in this work.

To our knowledge, Ref. \cite{Hamaus:2016} is the only previous work that performs a joint RSD-AP fit to the measured void-galaxy correlation. These authors use a different theoretical model for the RSD contribution to that used in this work (see Ref. \cite{Nadathur:2019a} for a comparison of the two models). In previous work using a single $N$-body simulation, this model was found to provide strongly biased reconstructions of both the growth rate and the Alcock-Paczynski ratio $\epsilon$ when applied to voids in both dark matter and galaxy mocks \cite{Hamaus:2015}. In particular \cite{Hamaus:2015} reports values of $\epsilon$ that differ at the $\gtrsim3\sigma$ level from $\epsilon=1$ expected for the simulation, leading the authors to conclude that systematic effects in their modelling dominate over the statistical errors for a CMASS-like sample. Applying this model to fit real data from the earlier CMASS DR11 catalogue, Ref. \cite{Hamaus:2016} finds $\beta=0.417\pm0.089$ and $\epsilon=1.003\pm0.012$ (statistical errors only).

\begin{figure}
\centering
\includegraphics[width=0.95\linewidth]{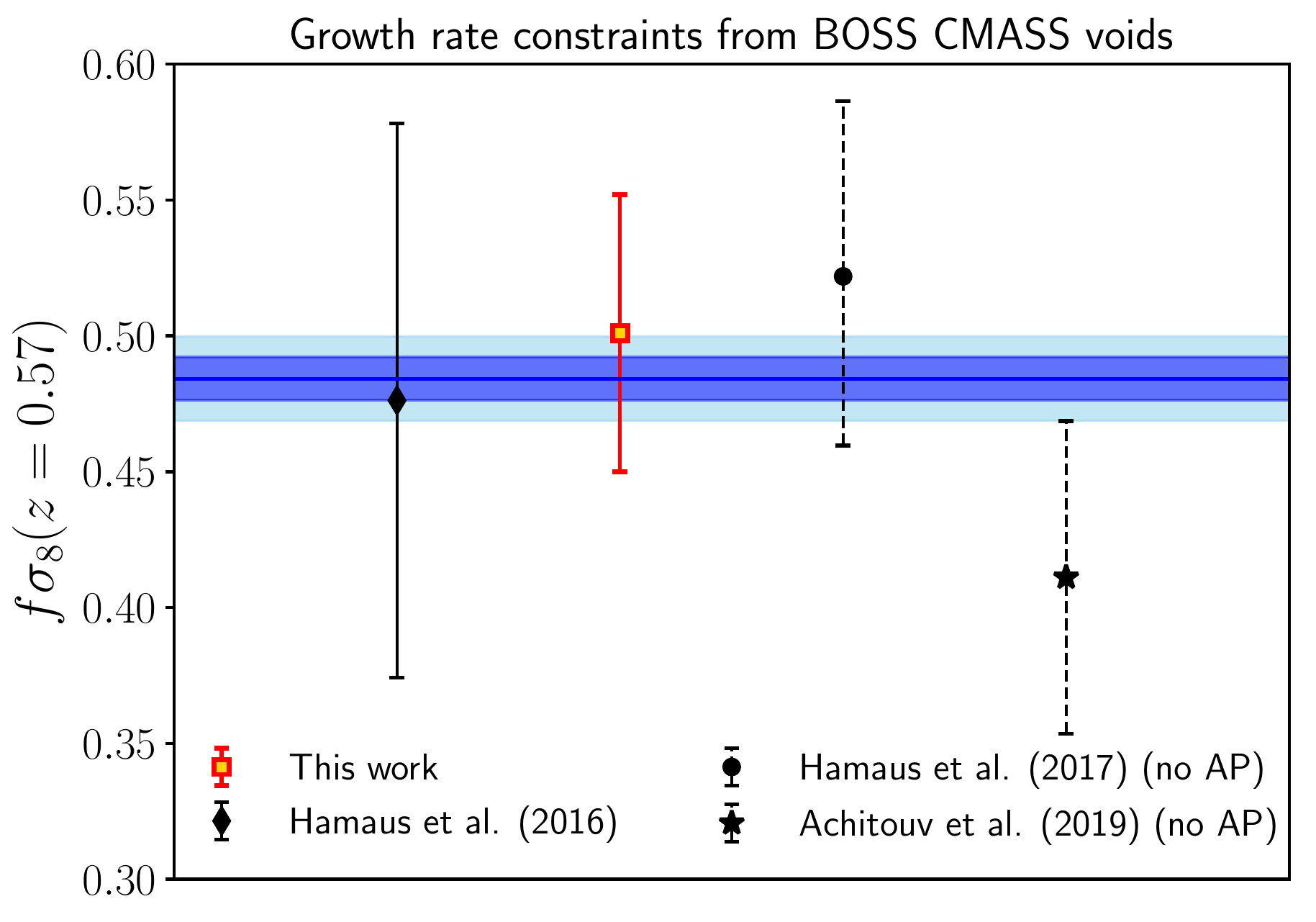}
\caption{Growth rate measurements and associated statistical uncertainties from different analyses of the anisotropic void-galaxy correlation in the BOSS CMASS dataset compared to the result in this work. The blue line and dark (light) blue bands show the mean and $68\%$ ($95\%$) confidence limits from Planck extrapolated to the CMASS redshift assuming a \lcdm ~model. For fair comparison, constraints on $\beta=f/b$ quoted in Refs. \cite{Hamaus:2016,Hamaus:2017a,Achitouv:2019} have been converted in terms of $f\sigma_8$ as described in the text. Values from Refs. \cite{Hamaus:2017a} and \cite{Achitouv:2019} are shown with dashed lines as these studies assume a fixed cosmology and do not simultaneously fit for Alcock-Paczynski distortions.}
\label{fig:voidfs8vals}
\end{figure}

We summarize the growth rate measurements from different analyses of the void-galaxy correlation in the CMASS data in Figure \ref{fig:voidfs8vals}. The growth rate results from Refs. \cite{Hamaus:2016,Hamaus:2017a,Achitouv:2019} are all given in terms of $\beta$, as quoted above. Our analysis provides a much tighter constraint on $\beta$ as a derived parameter, $\beta=0.37\pm0.01$. However as argued in Section \ref{subsec:fits} we regard $b$ as a nuisance parameter and thus report our headline results in terms of the quantity $f\sigma_8$. To provide a fair comparison we therefore translate the constraints on $\beta=f/b$ provided in Refs. \cite{Hamaus:2016,Hamaus:2017a,Achitouv:2019} in terms of $f\sigma_8$. To do this we take fiducial values $b=1.85$ \cite{Alam:2017} and $\sigma_8(z=0.57)=0.602$, the central value for the Planck \lcdm~cosmology \cite{Planck:2018params}, and assume both of these quantities are known exactly without error. The resultant $f\sigma_8$ constraints are $f\sigma_8=0.48\pm0.10$ (for \cite{Hamaus:2016}), $f\sigma_8=0.522^{+0.065}_{-0.062}$ (for \cite{Hamaus:2017a}) and $f\sigma_8=0.411\pm0.058$ (for \cite{Achitouv:2019}), and are shown in Figure \ref{fig:voidfs8vals}. The error bars for the points from Refs. \cite{Hamaus:2017a} and \cite{Achitouv:2019} are shown with dashed lines to emphasize that these are underestimates of the true uncertainty in these analyses as they do not allow for the additional Alcock-Paczynski contribution.

\section{Combined BOSS results}
\label{sec:combination}

\subsection{Comparison with BAO and RSD}
\label{subsec:comparison to BAO}

We now turn to a comparison of the results obtained from the void analysis presented here to those obtained from traditional galaxy clustering methods based on fitting for the BAO peak and RSD in the galaxy correlation. In addition to $f\sigma_8$, galaxy clustering analyses provide fits for the Alcock-Paczynski parameters defined to include the sound horizon at the drag epoch, $r_s$, which determines the location of the BAO peak:
\begin{equation}
    \label{eq:BAOalphas}
    \alpha_\perp = \frac{D_A(z)r_s^\mathrm{fid}}{D_A^\mathrm{fid}(z)r_s},\;\;\;\alpha_{||}=\frac{H^\mathrm{fid}(z)r_s^\mathrm{fid}}{H(z)r_s}\,,
\end{equation}
whereas our void analysis does not make reference to $r_s$ (equation \ref{eq:alphas}). Table \ref{table:constraints} 
\begin{table*}
\caption{\label{table:constraints} Mean values and $1\sigma$ errors for parameters measured from the BOSS DR12 CMASS data using the void-galaxy cross-correlation in this work (labelled `voids'), BAO \cite{Gil-Marin:2016a}, anisotropic galaxy clustering (RSD; \cite{Gil-Marin:2016b}) and combinations of methods.}
\begin{ruledtabular}
\begin{tabular}{@{}lccccc}
Parameter & voids & BAO & RSD & BAO + RSD & BAO + RSD + voids \\
\hline
$D_A(0.57)H(0.57)/c$ & $0.4367\pm 0.0045$ & $0.463\pm0.017$ & $0.437\pm0.018$ & $0.449\pm0.014$ & $0.4396\pm 0.0040$ \\
$f(0.57)\sigma_8(0.57)$ & $0.501\pm 0.051$ & $-$ & $0.444\pm0.038$ & $0.462\pm0.032$ & $0.453\pm 0.022$ \\
$D_A(0.57)/r_s$ & $-$ & $9.47\pm0.12$ & $9.42\pm0.15$ & $9.44\pm0.11$ & $9.383\pm 0.077$ \\
$H(0.57)r_s\;[10^3\,\mathrm{km\,s}^{-1}]$ & $-$ & $14.67\pm0.42$ & $13.92\pm0.44$ & $14.28\pm0.34$ & $14.05\pm 0.14$ \\
\end{tabular}
\end{ruledtabular}
\end{table*}
summarizes the constraints obtained on $F_\mathrm{AP}=D_AH/c$, $f\sigma_8$, $D_A/r_s$ and $Hr_s$ from using the void-galaxy correlation, BAO and galaxy RSD respectively, for the same BOSS DR12 CMASS data. Galaxy BAO results are taken from Ref. \cite{Gil-Marin:2016a} and RSD results from Ref. \cite{Gil-Marin:2016b}. As previously noted, the void analysis provides a much more precise measure of $F_\mathrm{AP}$ than can be obtained from galaxy clustering, but does not constrain $D_A/r_s$ or $Hr_s$ individually.

\begin{figure*}
\centering
\includegraphics[width=0.48\linewidth]{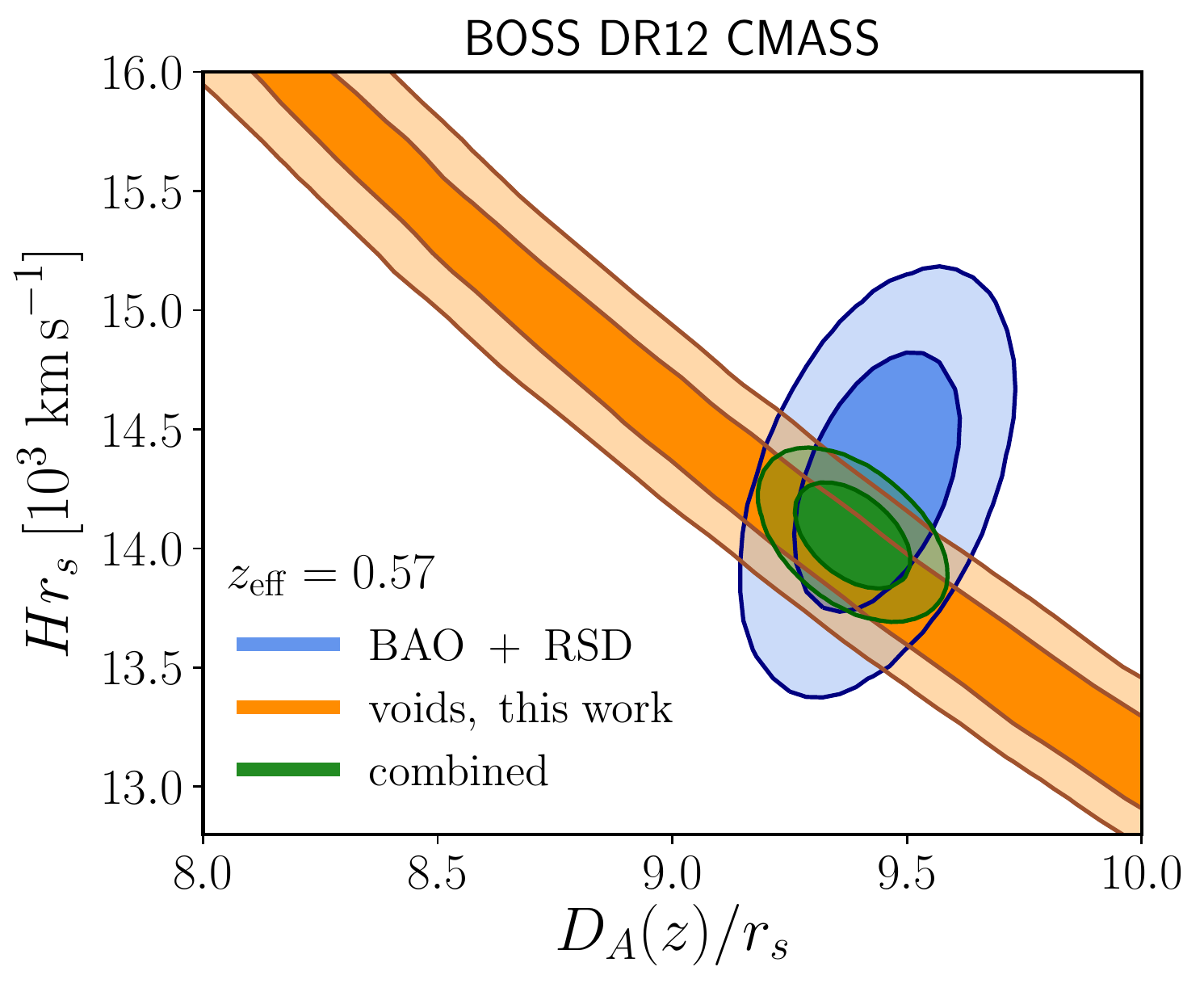}
\includegraphics[width=0.45\linewidth]{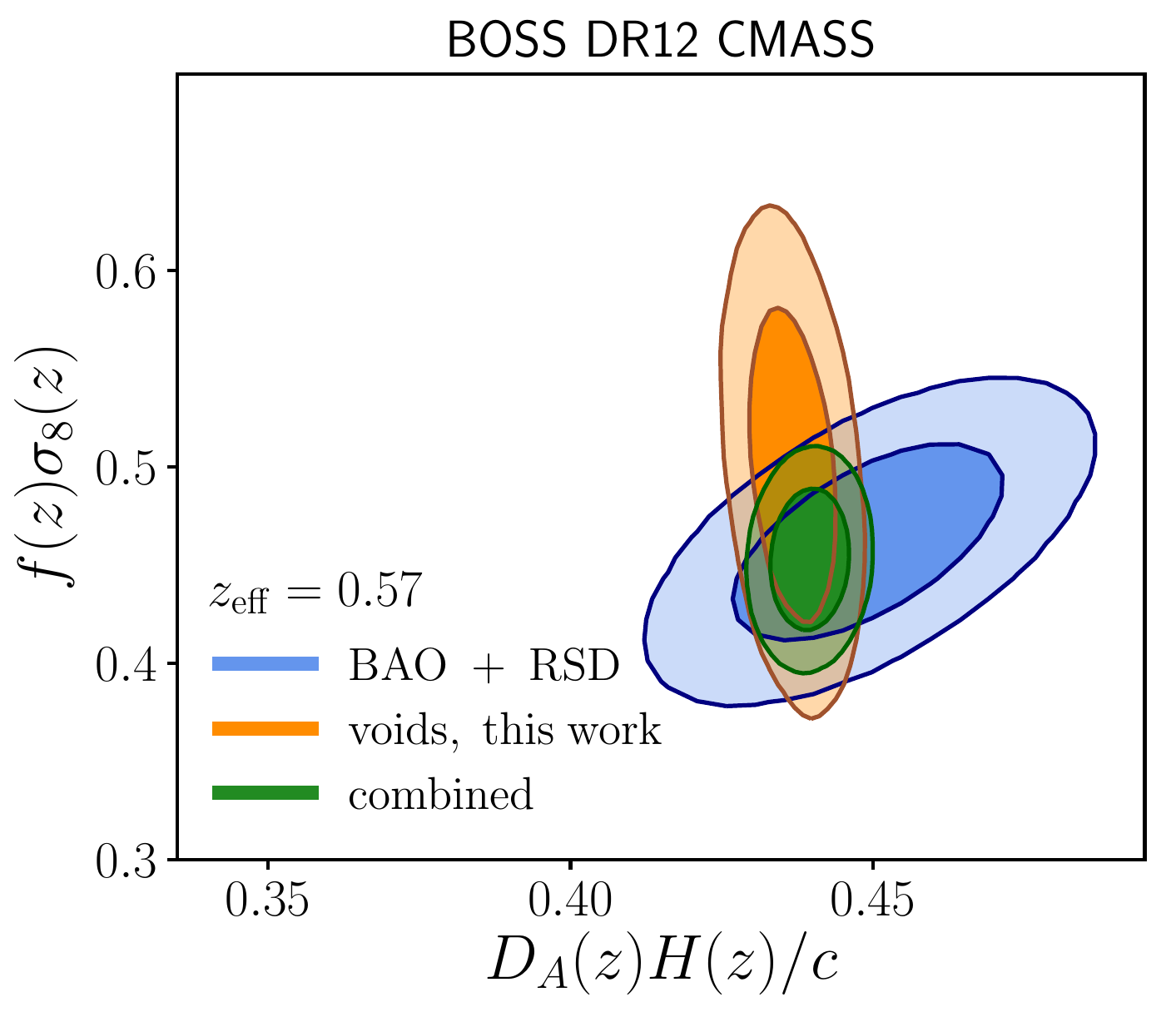}
\caption{Marginalized likelihood contours for parameters obtained from the BOSS DR12 CMASS catalogue using our measurement of the void-galaxy cross-correlation (in orange), compared to those from BAO and RSD galaxy clustering analyses \cite{Gil-Marin:2016a,Gil-Marin:2016b} (in blue). The methods are combined, accounting for their covariance as they use the same underlying survey data, to produce the joint likelihood constraints (in green). Contours enclose $68\%$ and $95\%$ of the total probability in each case.}
\label{fig:constraints}
\end{figure*}

To explore the consequences of this result, we plot the marginalized parameter posteriors from the different analyses in Figure \ref{fig:constraints}. Here BAO and galaxy RSD results from Refs. \cite{Gil-Marin:2016a} and \cite{Gil-Marin:2016b} have been combined together (blue contours) using the mock-based covariance method \cite{Sanchez:2017a} described in Section \ref{subsec:combination}. 

In the $D_A/r_s-Hr_s$ plane, the void analysis gives extremely tight constraints on statistical anisotropy in the $D_AH$ direction due to the precision achieved in measurement and modelling of the quadrupole $\xi^s_2$ at relatively small scales, and the fact that the Alcock-Paczynski effect and distortions due to galaxy outflow around voids make qualitatively different contributions to the quadrupole. This is also reflected in the fact that constraints on $D_AH$ from the void analysis are almost completely independent of those on $f\sigma_8$, in contrast to galaxy clustering results. $D_AH$ constraints from galaxy clustering are still driven primarily by measurement of the BAO peak location in the radial and transverse directions (in both pre-reconstruction and post-reconstruction analyses). This is because of the strong degeneracy in Alcock-Paczynski and RSD effects on the broadband part of the galaxy clustering at scales smaller than the BAO scale \cite{Ballinger:1996}. Measures of statistical isotropy using the radial and transverse BAO peak locations are less precise than for voids because of the much larger scales required, which restrict the sensitivity of a survey of finite volume. On the other hand, voids do not provide any constraint along the $D_V\propto (D_A^2/H)^{1/3}$ direction that is very well measured by BAO, due to the assumption that the absolute void size is not known independent of cosmology, so that measurements of the void-galaxy monopole cannot be used as an absolute ruler. 

It is clear from this figure that the void analysis is very complementary to galaxy clustering: the degeneracy directions in parameter space from the two methods are almost orthogonal, so a large information gain is available by combining them. 

\subsection{Combining measurements and likelihoods}
\label{subsec:combination}

To exploit the complementarity of the void-galaxy and galaxy clustering analyses, we need to combine them consistently while accounting for the covariance between the methods due to the fact that they are applied to the same underlying dataset. To do this, we employ the method used by the BOSS analysis \cite{Sanchez:2017a,Alam:2017} to obtain consensus constraints. 

For $m$ different analyses, each measuring $p$ parameters, we represent the results of the $i$th analysis by the $p$-dimensional vector $\mathbf{D}_i$ with $p\times p$ covariance matrix $\mathbf{C}_{ii}$, and the cross-covariance between the results of the $i$th and $j$th analyses by $\mathbf{C}_{ij}$. These can be combined to give a total $m\cdot p\,\times\,m\cdot p$ covariance matrix $\mathbf{C}_\mathrm{tot}$, and precision matrix $\mathbf{\Psi}_\mathrm{tot}=\mathbf{C}_\mathrm{tot}^{-1}$, which has corresponding block elements $\mathbf{\Psi}_{ij}$. For Gaussian posterior distributions, the consensus result obtained from combining the analyses is 
\begin{equation}
    \label{eq:consensus mean}
    \mathbf{D}_\mathrm{c} = \mathbf{\Psi}_\mathrm{c}^{-1}\sum_{i=1}^m \left(\sum_{j=1}^m \mathbf{\Psi}_{ji}\right)\mathbf{D}_i\,,
\end{equation}
with combined covariance matrix 
\begin{equation}
    \label{eq:consensus cov}
    \mathbf{C}_\mathrm{c} \equiv \mathbf{\Psi}_\mathrm{c}^{-1} \equiv \left(\sum_{i=1}^m \sum_{j=1}^m \mathbf{\Psi}_{ji}\right)^{-1}\,.
\end{equation}

When combining pre- and post-reconstruction galaxy clustering results (denoted BAO + RSD) we have $m=2$ and $p=3$ for parameters $f\sigma_8$, $D_A/r_s$ and $Hr_s$. As the post-reconstruction BAO analysis does not measure $f\sigma_8$, the uncertainty on this parameter from the BAO method is taken to be formally infinite in the relevant covariance matrix. BAO and RSD results for the CMASS sample are taken from Refs. \cite{Gil-Marin:2016a} and \cite{Gil-Marin:2016b}. To combine the galaxy clustering results with the void constraints (for the combination denoted BAO + RSD + voids), $m=3$ and we work in the parameter basis $\left(f\sigma_8, D_AH/c, D_V/r_s\right)$. In this case the uncertainty on $D_V/r_s$ from the void measurement is taken to be formally infinite. After combination, the consensus results and covariance matrix are translated back into the $\left(f\sigma_8, D_A/r_s, Hr_s\right)$ basis for easy comparison with previous works.

For the combination of results here, it is necessary to determine the cross-covariances $\mathbf{C}_{ij}$. Refs. \cite{Gil-Marin:2016a} and \cite{Gil-Marin:2016b} performed maximum likelihood fits for the galaxy clustering analyses for the same set of MD-Patchy mocks we use here.\footnote{This data was kindly provided to us by H\'ector Gil-Mar\'in.} These mock results can be combined with our own fits to the same mocks to determine the cross-covariance between the different methods.

\subsection{Joint results}
\label{subsec:joint}

The consensus mean values and marginalized $68\%$ confidence limits obtained from the combined analysis for CMASS are $f(0.57)\sigma_8(0.57) = 0.453\pm0.022$ (a 4.9\% measurement), $D_A(0.57)/r_s=9.383\pm0.077$ (0.8\%) and $H(0.57)r_s=(14.05\pm0.14)\times10^3\,\mathrm{km\,s}^{-1}$ (1\%), as summarized in Table \ref{table:constraints} and shown by the green contours in Figure \ref{fig:constraints}. These should be compared with the best BAO\dataplus RSD combination values $f(0.57)\sigma_8(0.57) = 0.462\pm0.032$, $D_A(0.57)/r_s=9.44\pm0.11$ and $H(0.57)r_s=(14.28\pm0.34)\times10^3\,\mathrm{km\,s}^{-1}$ without including voids. The addition of the void information therefore allows an improvement in measurement precision of $30$-$60\%$ on each parameter \emph{individually}. The new consensus mean values are all consistent with the previous BAO\dataplus RSD combination results to within the previous $1\sigma$ error bars.

\begin{figure}
\centering
\includegraphics[width=0.95\linewidth]{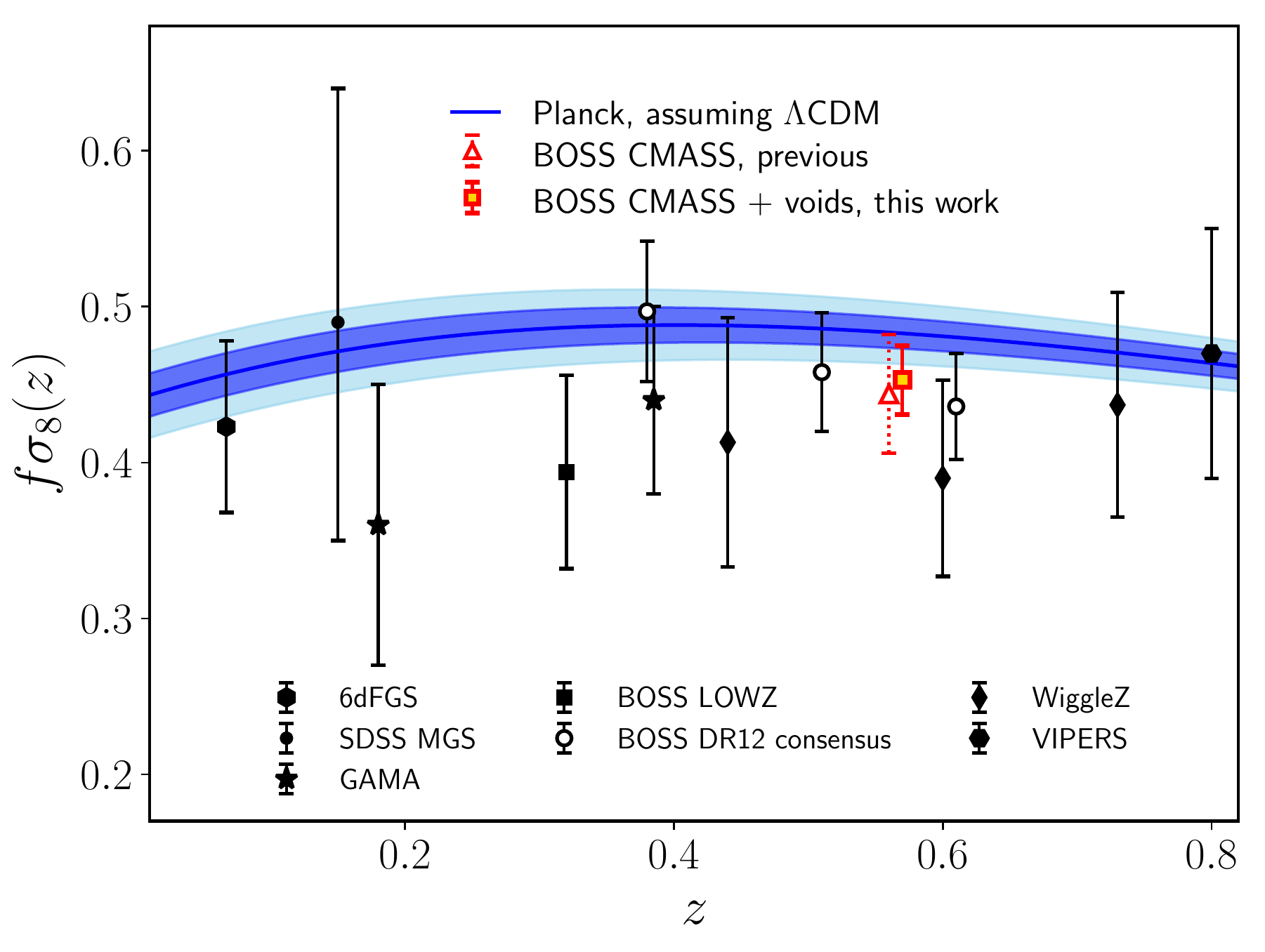}
\caption{The $f\sigma_8(z)$ value obtained in this work from BOSS DR12 CMASS, compared with values obtained from the 6dFGS \cite{Beutler:2012}, GAMA \cite{Blake:2013}, WiggleZ \cite{Blake:2012} and VIPERS \cite{Guzzo:2008,delaTorre:2013} surveys, as well as from the SDSS Main Galaxy Sample (MGS) \cite{Howlett:2015}, and BOSS DR12 analyses. For DR12, we show results from the LOWZ and CMASS samples \cite{Gil-Marin:2016b} and the final consensus results from the combined sample in three redshift bins \cite{Alam:2017} separately. The previous result from the same DR12 CMASS catalogue as used in this work is highlighted and shown slightly shifted from the effective redshift $z=0.57$ for clarity. The line and shaded contours indicate the value and constraints on $f\sigma_8$ obtained by extrapolating the CMB results from Planck to these redshifts assuming a $\Lambda$CDM cosmological model.}
\label{fig:fs8vals}
\end{figure}

Figure \ref{fig:fs8vals} shows the final joint $f\sigma_8$ result we obtain from CMASS compared to results from other galaxy surveys, including the previous best CMASS constraint from Ref. \cite{Gil-Marin:2016b}, and the Planck values extrapolated to low redshifts assuming a \lcdm~cosmology. The 4.9\% constraint we find is the tightest obtained from any galaxy survey to date.

Table \ref{table:CMASS covmat} 
\begin{table}[b]
\caption{\label{table:CMASS covmat}Consensus mean values and covariance matrix for parameters measured from the combined BAO + RSD + voids analysis for the BOSS DR12 CMASS galaxy sample at effective redshift $z=0.57$. The covariance matrix is symmetric, so lower triangle entries are omitted.}
\begin{ruledtabular}
\begin{tabular}{@{}lcccc}
Parameter & Mean &  \multicolumn{3}{c}{$10^4C_{ij}$}  \\
\colrule
$D_A(0.57)/r_s$ & $9.383$ & 60.031 & -56.265 & 3.3545 \\
$H(0.57)r_s\;[10^3\,\mathrm{km\,s}^{-1}]$ & $14.05$ & $-$ & 198.68 & -2.4322 \\
$f(0.57)\sigma_8(0.57)$ & $0.453$ & $-$ & $-$ & 4.9283  \\
\end{tabular}
\end{ruledtabular}
\end{table}
provides the full consensus covariance matrix for all three measured parameters from CMASS, for use in cosmological analyses. In this paper we have only analysed CMASS voids, for reasons explained above. Nevertheless, for completeness we use the likelihood combination method to also combine previously published galaxy BAO and RSD results for the BOSS LOWZ sample from Refs. \cite{Gil-Marin:2016a} and \cite{Gil-Marin:2016b}. The consensus means and covariance matrix for LOWZ at redshift $z=0.32$ are given in Table \ref{table:LOWZ covmat}.
\begin{table}[b]
\caption{\label{table:LOWZ covmat}Consensus mean values and covariance matrix for parameters measured from the combined BAO\dataplus RSD analysis for the BOSS DR12 LOWZ galaxy sample at effective redshift $z=0.32$.}
\begin{ruledtabular}
\begin{tabular}{@{}lcccc}
Parameter & Mean & \multicolumn{3}{c}{$10^3C_{ij}$}  \\
\colrule
$D_A(0.32)/r_s$ & 6.55 & 15.685 & 28.301 & 3.6319  \\
$H(0.32)r_s\;[10^3\,\mathrm{km\,s}^{-1}]$ & 11.57 & $-$ & 221.89 & 16.381 \\
$f(0.32)\sigma_8(0.32)$ & 0.428 & $-$ & $-$ & 2.995 \\
\end{tabular}
\end{ruledtabular}
\end{table}
The CMASS and LOWZ samples do not overlap in redshift and so the results in Tables \ref{table:CMASS covmat} and \ref{table:LOWZ covmat} are independent of each other. They can be compared with the consensus results for the combined LOWZ+CMASS sample in three overlapping redshift bins provided by the final BOSS DR12 analysis \cite{Alam:2017}, which has so far been regarded as providing the best BOSS constraints. In the next section, we compare the constraining power of these two approaches.

\section{Cosmological model parameters}
\label{sec:cosmology}

\subsection{Data sets}
\label{subsec:cosmo data}

We now turn to the cosmological interpretation of our results. In this section, we consider the final consensus CMASS results at $z=0.57$ from the combination of galaxy clustering and voids summarized in Table \ref{table:CMASS covmat}, which we label `CMASS BAO\dataplus RSD\dataplus voids' as well as the LOWZ results at $z=0.32$ from Table \ref{table:LOWZ covmat}, which are referred to as `LOWZ BAO\dataplus RSD'. When both of these are used together, we refer to the combination as `LOWZ\dataplus CMASS\dataplus voids'. This combination is compared to the BOSS DR12 consensus results in three redshift bins, $z=0.38$, $0.51$ and $0.61$, which are provided in Table 8 of Ref. \cite{Alam:2017}, referred to as `BOSS consensus'. The BOSS consensus results have already been shown to improve upon the LOWZ\dataplus CMASS constraints in two redshift bins when void information is not included \cite{Alam:2017}, so we do not consider this final combination here.

In all cases, we combine the BOSS results with CMB anisotropy data from the Planck 2015 results \cite{Planck:2015params}. To do this, we make use of temperature and polarization data at both low and high multipoles through the use of the public likelihoods \nolinkurl{plik\_dx11dr2\_HM\_v18\_TTTEEE} (for high multipoles) and \nolinkurl{lowl_SMW_70_dx11d_2014_10_03_v5c_Ap} (for low multipoles) from the 2015 Planck data release, but we do not include CMB lensing information. We note that the Planck 2015 results have since been updated by the 2018 data release \cite{Planck:2018params}, but at the time of preparing this paper the new likelihoods were not yet publicly available. The 2015 Planck data is the same as used in the BOSS consensus paper \cite{Alam:2017}, which aids direct comparison.

The point of the analysis in this section is to demonstrate the information gain from the use of voids relative to the consensus BOSS results, rather than to update the current best available cosmological constraints (which would in any case require the Planck 2018 likelihoods). We therefore do not include BAO or RSD information from any other galaxy or quasar surveys, or measurements of the distance-redshift relation from Type Ia supernovae, but our analysis can easily to be extended to include these cases.

We use the above combinations of results to constrain parametrized cosmological models using MCMC chains run using the standard cosmological package CosmoMC \cite{Lewis:2002,Lewis:2013}, trivially modified to include the covariances in Tables \ref{table:CMASS covmat} and \ref{table:LOWZ covmat} where appropriate.

\subsection{\lcdm ~model}
\label{subsec:lcdm}

\begin{figure}
\centering
\includegraphics[width=0.95\linewidth]{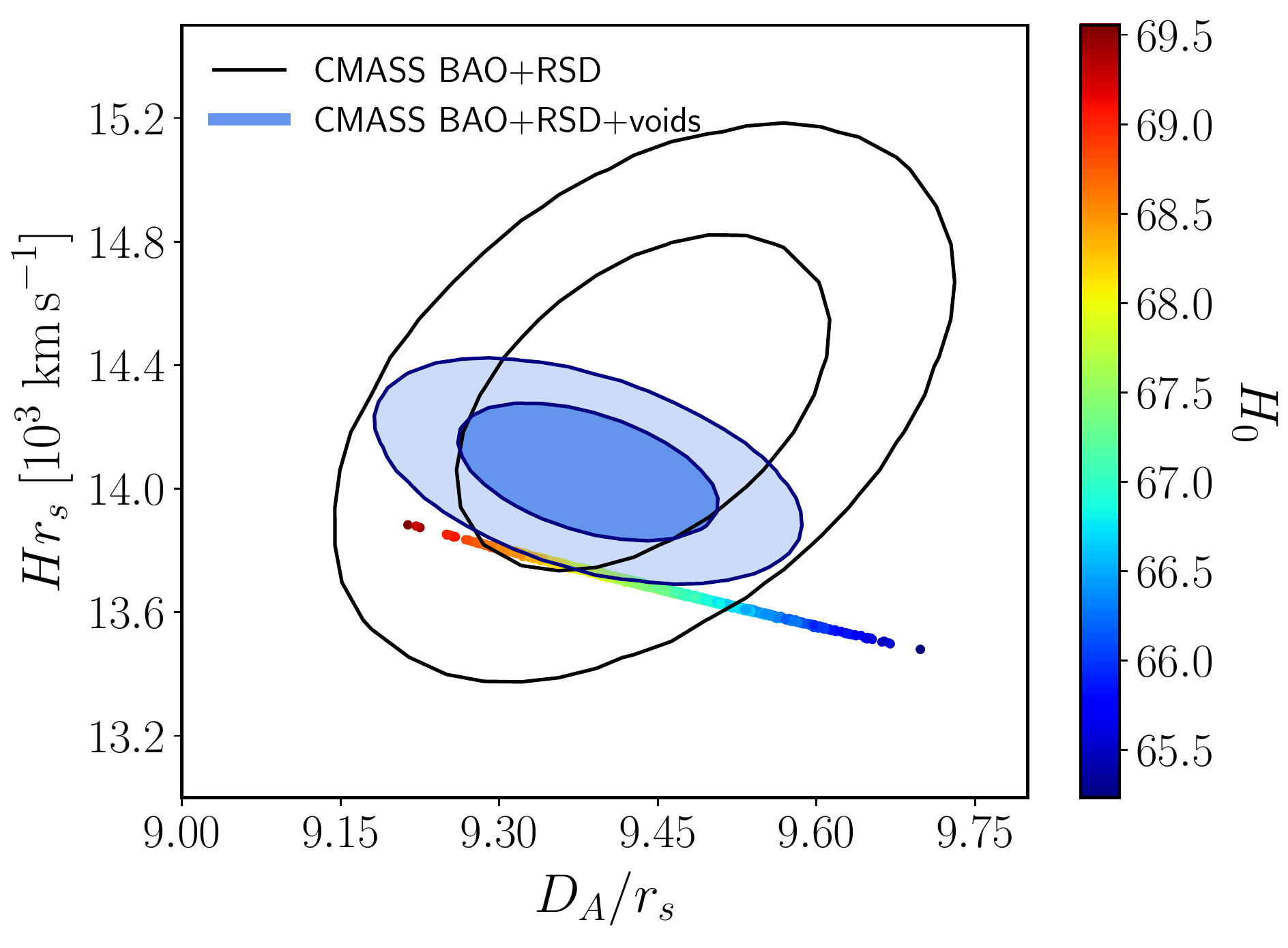}
\caption{Likelihood contours showing the $68\%$ and $95\%$ confidence intervals for $D_A/r_s$ and $Hr_s$ at the CMASS effective redshift $z=0.57$ from BAO and galaxy RSD (black lines) and from BAO, RSD and the additional void-galaxy information in this work (blue contours). The points show samples from fits of the \lcdm~model to Planck CMB data, coloured by the $H_0$ value. The CMASS results are consistent with each other as well as with Planck. Note that the addition of void data has changed the degeneracy direction of the likelihood contours.}
\label{fig:Planck}
\end{figure}

We first consider the impact of the improved precision of the parameter measurements from CMASS void data within the context of a flat \lcdm~cosmological model. As can be seen from Table \ref{table:constraints}, although the addition of void data greatly tightens the constraints on $f\sigma_8$, $D_A/r_s$ and $Hr_s$, it does not lead to significant shifts in the central values and therefore does not change the overall consistency of BOSS with CMB constraints from Planck within \lcdm. This is represented in Figure \ref{fig:Planck}, which shows the CMASS constraints in the $D_A/r_s-Hr_s$ plane together with samples from the \lcdm~model chains fit to Planck data alone, coloured by the value of $H_0$. While the error ellipses have significantly shrunk, the CMASS central values remain somewhat shifted from the Planck results, in particular the value of $H(0.57)r_s$ is slightly high. This small difference is driven by the post-reconstruction BAO results \cite{Gil-Marin:2016a} rather than either the void results in this work or those from the pre-reconstruction full-shape RSD fits \cite{Gil-Marin:2016b}, which overlap the Planck constraints at the $1\sigma$ level both individually and in combination. This can also be seen in Table \ref{table:constraints}, where the BAO results are responsible for pulling $H(0.57)r_s$ to higher values.

The addition of CMASS void data does not significantly shift the central parameter values obtained within \lcdm, but does lead to a reduction of the final error bars. We quantify this in a simple way by calculating the total volume of the likelihood region in $N$-dimensional parameter space, 
\begin{equation}
    \label{eq:vol}
    V = (\mathrm{det}\,\mathbf{C})^{-1/2}\,,
\end{equation}
where $\mathbf{C}$ is the covariance matrix describing the posterior. Comparing the results for the Planck\dataplus BOSS consensus and Planck\dataplus LOWZ\dataplus CMASS\dataplus voids combination described above, we find that the addition of void information reduces the allowed volume of parameter space by 11\% when considering the 6 main cosmological parameters of base \lcdm. 

The Planck\dataplus BOSS consensus data combination constrains the Hubble constant $H_0$ and matter density $\Omega_m$ to be 
\twoonesig{H_0 &= (67.62\pm 0.48)\;\mathrm{km\,s}^{-1}\,\mathrm{Mpc}^{-1},}{\Omega_m &= 0.3105\pm 0.0064\,.}
{\mksym{Planck}\dataplus\\\mksym{BOSS}\,\mksym{consensus}   \label{eq:lcdm consensus}}
With the addition of the CMASS void data these constraints are strengthened to  
\twoonesig{H_0 &= 67.71\pm 0.43\;\mathrm{km\,s}^{-1}\,\mathrm{Mpc}^{-1},}{\Omega_m &= 0.3093\pm 0.0057\,,}
{\mksym{Planck}\dataplus\\\mksym{LOWZ}\dataplus\mksym{CMASS}\\\dataplus\mksym{voids}   \label{eq:lcdm voids}}
representing an $\sim11\%$ reduction in the uncertainty on both $H_0$ and $\Omega_m$. Thus the addition of void information to LOWZ, CMASS and the 2015 Planck data gives constraints comparable to those from the latest 2018 Planck release and the BOSS consensus results when \emph{also} including CMB lensing and BAO data from several additional galaxy surveys at different redshifts \cite{Planck:2018params}.

\begin{figure*}
\centering
\includegraphics[width=0.331\linewidth]{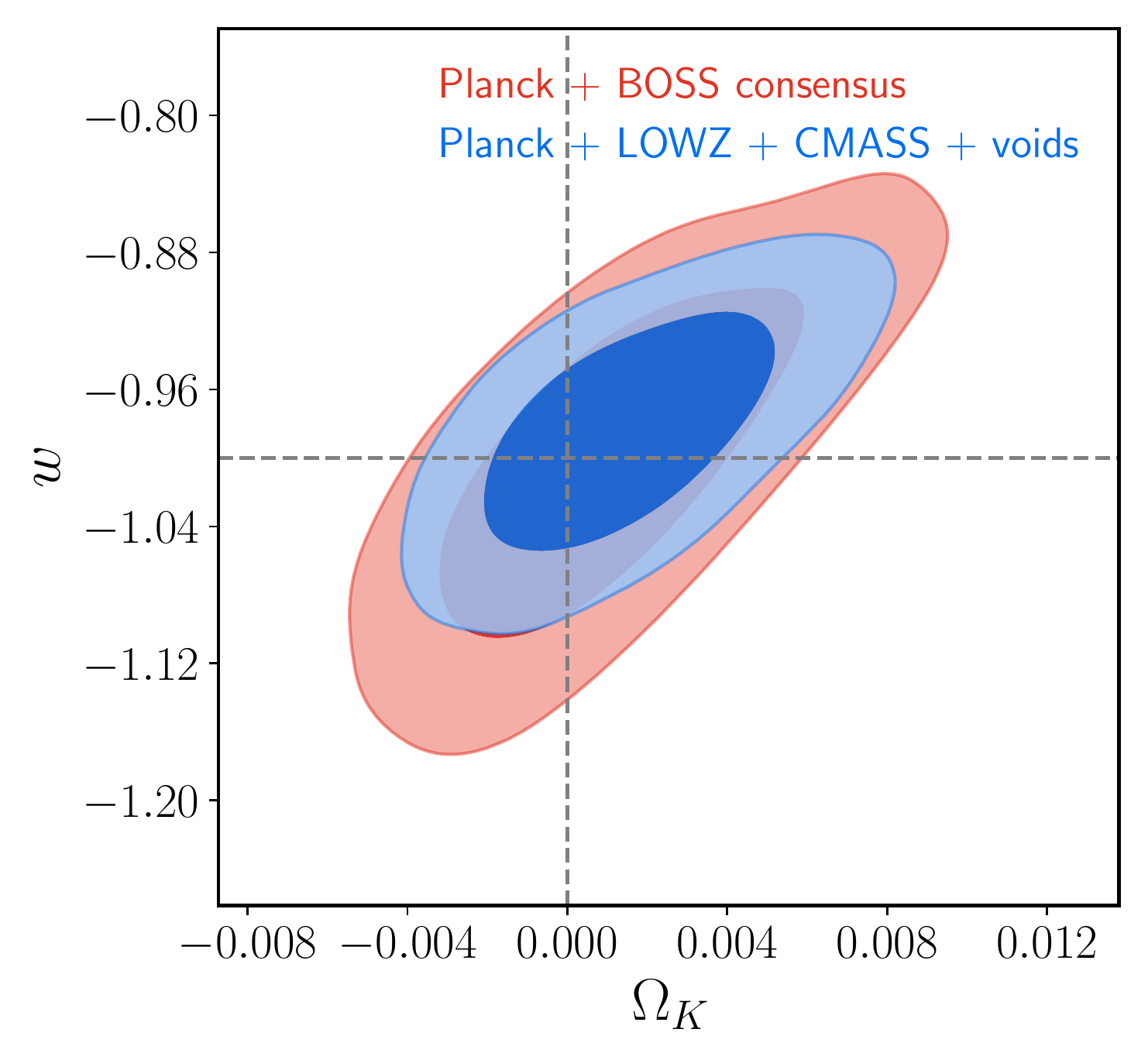}
\includegraphics[width=0.325\linewidth]{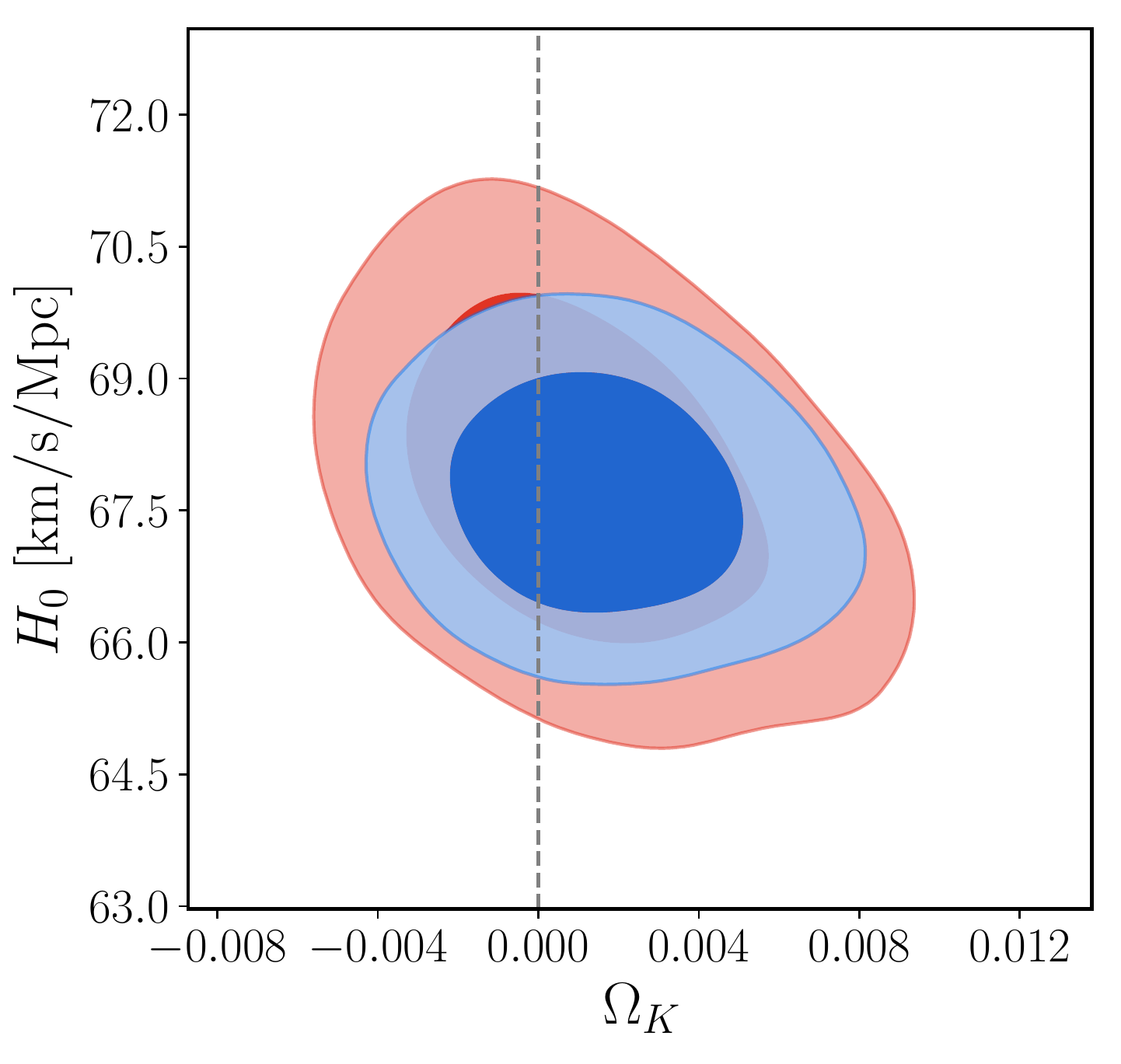}
\includegraphics[width=0.325\linewidth]{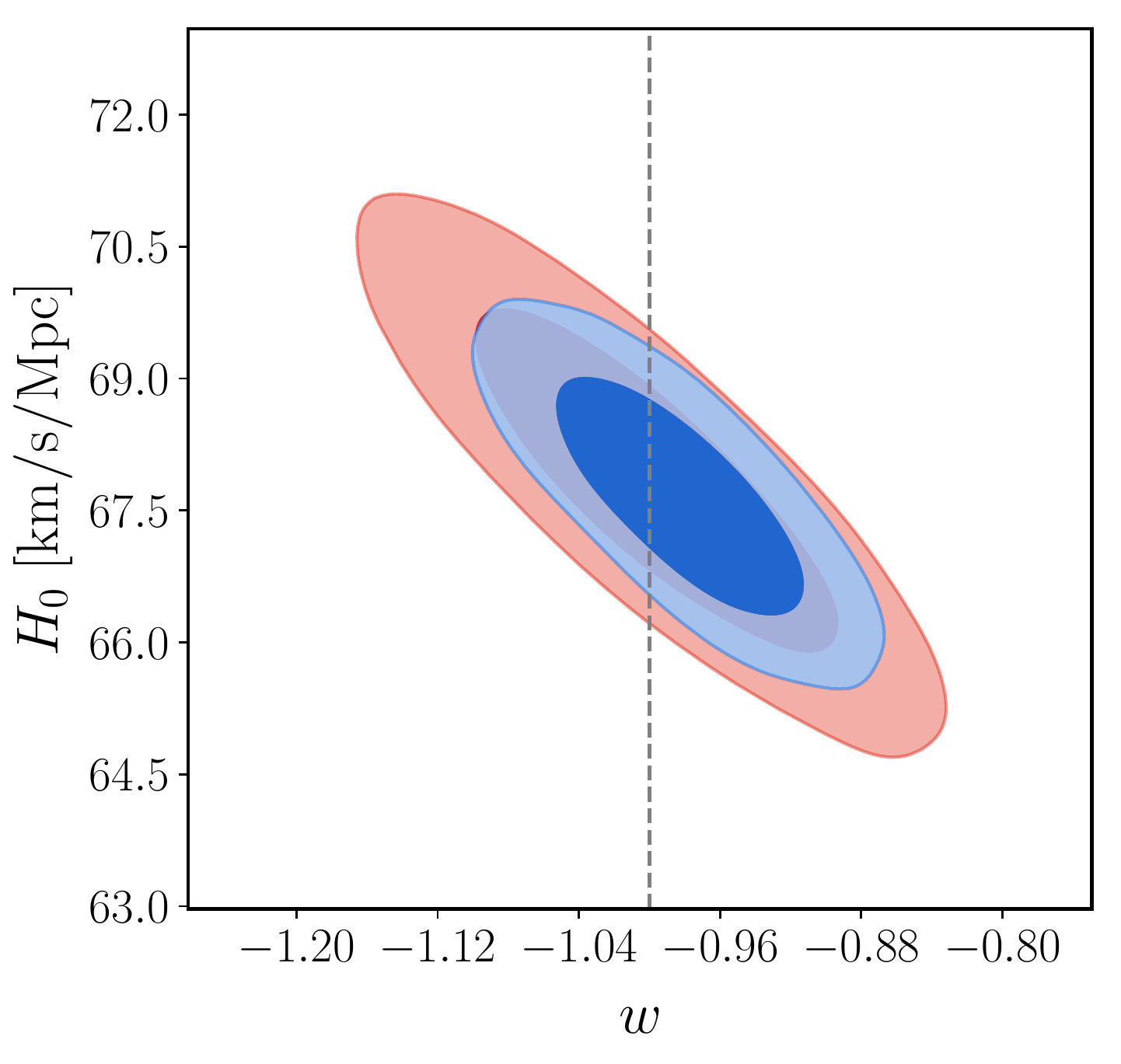}
\caption{Parameter constraints on the dark energy equation of state $w$, curvature $\Omega_K$ and Hubble constant $H_0$ in 2-parameter extension to the standard flat $\Lambda$CDM cosmological model referred to as $ow$CDM, comparing the results for the combination of Planck with BOSS consensus results \cite{Alam:2017} (red contours), and with the addition of the CMASS void results from this work (blue). Contours enclose $68\%$ and $95\%$ of the probability. The improved Alcock-Paczynski measurement provided by our void analysis significantly tightens constraints on dark energy and curvature in this class of models without the need for additional primary data.}
\label{fig:owCDM}
\end{figure*}

\subsection{Dark energy and curvature}
\label{subsec:owcdm}

If a base-\lcdm~cosmological model is assumed,  Figure \ref{fig:Planck} shows that the extrapolation of Planck CMB results alone down to redshift $z=0.57$ already quite tightly constrains $D_A/r_s$ and $Hr_s$, particularly along the direction of the Alcock-Paczynski parameter combination $F_\mathrm{AP}=D_A(z)H(z)/c$. Although low-redshift large-scale structure measurements add some information in this scenario, their additional constraining power is much greater for models beyond \lcdm, which have parameter degeneracies that cannot be broken by CMB temperature and polarization data alone.\footnote{Although measurement of gravitational lensing of the CMB \cite{Planck:2018lensing} also helps break these degeneracies.} The same is true with the addition of the void constraints presented here.

To demonstrate the power of the void analysis for such models, we consider an example two-parameter extension of the base-\lcdm~model which allows for a dark energy equation of state parameter $w$ that can differ from $-1$, and deviations from flat geometry parametrized by curvature parameter $\Omega_K$. Following Ref. \cite{Alam:2017}, we refer to this as the $ow$CDM model. This extension allows the freedom to change the low-redshift behaviour while still keeping the angular diameter distance to recombination fixed, a degeneracy that is broken by the addition of distance-scale measurements at $z<1$. 

We fit the class of $ow$CDM models to the Planck\dataplus BOSS consensus and Planck\dataplus LOWZ\dataplus CMASS\dataplus voids combinations of datasets separately. The much greater precision in the measurement of the Alcock-Paczynski parameter at $z=0.57$ provided by the addition of void data breaks the low-redshift degeneracy and thus significantly tightens constraints compared to the Planck\dataplus BOSS consensus results: by the measure in equation \ref{eq:vol} the total allowed volume in parameter space shrinks by as much as $47\%$ when including voids. This can be seen in the significant shrinking of the marginalized 2D likelihood contours for this model shown in Figure \ref{fig:owCDM}. The marginalized constraints on the parameters of the model change from
\threeonesig{\Omega_K &=  0.0012^{+0.0025}_{-0.0033}\,,}{w &= -0.999\pm 0.068,}{H_0 &= 67.9\pm 1.3 \;\mathrm{km\,s}^{-1}\,\mathrm{Mpc}^{-1},}{\mksym{Planck}\dataplus\mksym{BOSS\,consensus}   \label{eq:owcdm consensus}}
to 
\threeonesig{\Omega_K &=  0.0015^{+0.0022}_{-0.0026}\,,}{w &= -0.983\pm 0.047,}{H_0 &= 67.67\pm 0.90\;\mathrm{km\,s}^{-1}\,\mathrm{Mpc}^{-1}.}{\mksym{Planck}\dataplus\\\mksym{LOWZ}\dataplus\mksym{CMASS}\\\dataplus\mksym{voids}   \label{eq:owcdm voids}}
These values are statistically consistent with the expectations $\Omega_K=0$, $w=-1$ for the base-\lcdm~model. The addition of void constraints greatly reduces the parameter uncertainty with respect to those from Planck\dataplus BOSS consensus (especially on $w$ and $H_0$), but does not lead to a significant shift of the central values. 

\section{Conclusion}
\label{sec:conclusion}

We have presented a measurement of the anisotropic void-galaxy cross-correlation in the BOSS DR12 CMASS data sample and a multipole-based analysis that jointly fits for distortions produced by the peculiar velocities of galaxies around voids and Alcock-Paczynski distortions due to deviations from the assumed cosmological model parameters. Our work uses the improved theoretical model for RSD effects introduced in Ref. \cite{Nadathur:2019a}, that provides an accurate description of $N$-body simulation results on all scales. This is the first void analysis of galaxy survey data that incorporates the use of velocity-field reconstruction to remove the complicating effect of RSD in the void centre positions themselves \cite{Nadathur:2019b}. These void centre RSD effects are extremely important in determining the observed multipoles, but their effect is not modelled in this or any other model of the void-galaxy correlation, so the reconstruction step is crucial to enable a consistent comparison of data and theory.

Our void analysis provides an extremely tight constraint on the Alcock-Paczynski parameter $F_\mathrm{AP}=D_A(z)H(z)/c = 0.4367\pm0.0045$ at redshift $z=0.57$, a $\sim1\%$ measure of the ratio of distances perpendicular to and along the line-of-sight direction and a factor of $\sim3.5$ more precise than the equivalent result from fitting the BAO peak position observed in the same CMASS data \cite{Gil-Marin:2016a}. This measurement has been validated through tests on 1000 MD-Patchy mock catalogues, which show that the systematic error in our measurement of $F_\mathrm{AP}$ is negligible. Our analysis also provides a $10\%$ constraint on the growth rate, $f\sigma_8(z=0.57)=0.501\pm0.051$, with no significant degeneracy between $f\sigma_8$ and $F_\mathrm{AP}$. This is significantly tighter than the $\sim21\%$ growth rate constraint previously reported from a joint RSD-AP fit to the void-galaxy correlation using CMASS data from the earlier DR11 release \cite{Hamaus:2016}. Our growth constraints are also better than those from various other void-galaxy analyses \cite{Achitouv:2017a, Hamaus:2017a, Hawken:2017, Achitouv:2019} which fit only for RSD effects at a fixed fiducial cosmology.

We have also for the first time demonstrated how the results of such a void analysis can be combined with traditional galaxy clustering measurements of the pre-reconstruction galaxy power spectrum and post-reconstruction BAO peak location to provide a large information gain. Combining with BAO and galaxy RSD results for the same DR12 CMASS galaxy sample from \cite{Gil-Marin:2016a,Gil-Marin:2016b}, we report an improvement in precision of $30$-$60\%$ in the measurement of each of the main parameters $f\sigma_8$, $D_A/r_s$, and $Hr_s$ individually. This very significant improvement, equivalent to the gain from a hypothetical $\sim300\%$ increase in the effective BOSS survey volume if using traditional galaxy clustering methods, is achieved using only the existing spectroscopic galaxy sample and without the need for any additional primary data. This in turn leads to important improvement in the constraints on cosmological model parameters over those obtained using the final BOSS DR12 consensus results \cite{Alam:2017}, both in the base 6-parameter \lcdm~model and in an extension that allows for a non-flat geometry and equation of state parameter $w\neq-1$.

This paper represents a concrete realization applied to real survey data of the promise of void-based measurements \cite{Lavaux:2012}. The information gain from the inclusion of the void-galaxy results is large enough that the methods outlined here should be considered an essential part of the toolkit for large-scale structure analysis for any spectroscopic galaxy survey, and in particular for the data to be generated in the near future from DESI \cite{2016arXiv161100036D} and Euclid \cite{2014IAUS..306..375S}. 

\begin{acknowledgments}
We wish to thank H\'ector Gil-Mar\'in for providing data on the likelihood fits to MD-Patchy mocks for the BAO and RSD analyses, and James Rich for comments on an earlier version of this paper that led to improvements in the analysis. S.N. acknowledges useful discussions with Florian Beutler. This research was supported by UK Space Agency grant ST/N00180X/1, and by the European Research Council through grant 646702 (CosTesGrav). Computational work was performed on the UK {\small SCIAMA} High Performance Computing cluster supported by the ICG, SEPNet and the University of Portsmouth.

This work made use of public catalogues from the SDSS-III BOSS Data Release 12. Funding for SDSS-III has been provided by the Alfred P. Sloan Foundation, the Participating Institutions, the National Science Foundation, and the U.S. Department of Energy Office of Science. The SDSS-III web site is \url{http://www.sdss3.org/}.

SDSS-III is managed by the Astrophysical Research Consortium for the Participating Institutions of the SDSS-III Collaboration including the University of Arizona, the Brazilian Participation Group, Brookhaven National Laboratory, Carnegie Mellon University, University of Florida, the French Participation Group, the German Participation Group, Harvard University, the Instituto de Astrofisica de Canarias, the Michigan State/Notre Dame/JINA Participation Group, Johns Hopkins University, Lawrence Berkeley National Laboratory, Max Planck Institute for Astrophysics, Max Planck Institute for Extraterrestrial Physics, New Mexico State University, New York University, Ohio State University, Pennsylvania State University, University of Portsmouth, Princeton University, the Spanish Participation Group, University of Tokyo, University of Utah, Vanderbilt University, University of Virginia, University of Washington, and Yale University.
\end{acknowledgments}

\emph{Author contributions}: S.N. and W.J.P. conceived the idea and developed the theoretical modelling. S.N. designed the methodology, led the data analysis, developed the code and drafted the paper. P.M.C. and H.A.W developed the code, contributed to the data analysis, and performed the MCMC for cosmological models. J.E.B. contributed to code development. S.N., P.M.C., W.J.P. and H.A.W performed the scientific interpretation of results. All authors contributed to editing the paper. 

\bibliography{refs}
\bibliographystyle{mod-apsrev4-2.bst}

\end{document}